\documentclass[final,3p,times,twocolumn]{elsarticle}

\usepackage{amssymb}
\usepackage{amsmath, xparse}
\usepackage{array}
\usepackage{graphicx}
\usepackage[table,xcdraw]{xcolor}
\usepackage[export]{adjustbox}
\biboptions{numbers,sort&compress}

\newcommand{\etal}{{et al.}}

\newenvironment{conditions}
  {\par\vspace{\abovedisplayskip}\noindent\begin{tabular}{>{$}l<{$} @{${}={}$} l}}
  {\end{tabular}\par\vspace{\belowdisplayskip}}

\journal{Artificial Intelligence in Medicine}

\begin{document}
\graphicspath{{images/}}

\begin{frontmatter}

\title{CapillaryNet: An Automated System to Quantify Skin Capillary Density and Red Blood Cell Velocity from Handheld Vital Microscopy}

\author[inst1]{Maged Abdalla Helmy}
\author[inst2]{Tuyen Trung Truong}
\author[inst3]{Anastasiya Dykky}
\author[inst1]{Paulo Ferreira}
\author[inst1]{Eric Jul}

\affiliation[inst1]{organization={Department of Informatics, University of Oslo},
            addressline={Gaustadalleen 21}, 
            email={magedaa,paulofe,ericbj@uio.no},
            city={Oslo},
            postcode={0373}, 
            country={Norway}}

\affiliation[inst2]{organization={Department of Mathematics, University of Oslo},
            addressline={Gaustadalleen 21}, 
            email={tuyentt@math.uio.no},
            city={Oslo},
            postcode={0373}, 
            country={Norway}}

\affiliation[inst3]{organization={ODI Medical AS},
            addressline={Gaustadalleen 23},
            email={anastasiya.dykyy@odimedical.com},
            city={Oslo},
            postcode={0373}, 
            country={Norway}}
            
\begin{abstract}

Capillaries are the smallest vessels in the body responsible for delivering oxygen and nutrients to surrounding cells. Various life-threatening diseases are known to alter the density of healthy capillaries and the flow velocity of erythrocytes within the capillaries.
In previous studies, capillary density and flow velocity were manually assessed by trained specialists.
However, manual analysis of a standard 20-second microvascular video requires 20 minutes on average and necessitates extensive training.
Thus, manual analysis has been reported to hinder the application of microvascular microscopy in a clinical environment.
To address this problem, this paper presents a fully automated state-of-the-art system to quantify skin nutritive capillary density and red blood cell velocity captured by handheld-based microscopy videos.
The proposed method combines the speed of traditional computer vision algorithms with the accuracy of convolutional neural networks to enable clinical capillary analysis.
The results show that the proposed system fully automates capillary detection with an accuracy exceeding that of trained analysts and measures several novel microvascular parameters that had eluded quantification thus far, namely, capillary hematocrit and intracapillary flow velocity heterogeneity.
The proposed end-to-end system, named CapillaryNet, can detect capillaries at $\sim$0.9 seconds per frame with $\sim$93\% accuracy.
The system is currently being used as a clinical research product in a larger e-health application to analyse capillary data captured from patients suffering from COVID-19, pancreatitis, and acute heart diseases.
CapillaryNet narrows the gap between the analysis of microcirculation images in a clinical environment and state-of-the-art systems.

\end{abstract}


\end{frontmatter}

\section{Introduction}
\label{intro}
\thispagestyle{empty}

Capillaries are the smallest blood vessels in the body, measuring less than 20 micrometers in diameter \cite{De_Backer2007-hx}. 
In capillaries, several nutrients, including water, oxygen, and other nutrients essential to the maintenance of cellular metabolism, are exchanged with the interstitial fluid \cite{arefa2020,Shore2000-mu}.
A network of capillaries constitutes the microcirculation architecture of the body \cite{Guven2020-my}. 
Studies on the assessment of microcirculation have revealed that various diseases alter the capillary density and flow velocity in patients \cite{JS_Parker2019-vp, Nama2011-cz, Houben2017-bz, De_Graaff2003-rv, Fagrell1997-ht, Houtman1985-mo,Schmeling2011-jo, Duscha1999-zu,Robbins2011-df,Moeini2018-tc,Lopez2015-se, De_Backer2002-ae,Wester2014-tz}. 
Furthermore, in some studies, microcirculation has been claimed to alter during an early stage of disease progression; thus, microcirculation monitoring may be used for early detection of various clinical conditions \cite{Duscha1999-zu,Wester2014-tz}. 
The density of capillaries is an indicator of the surface area available for the exchange of nutrients  \cite{Shore2000-mu}. 
The capillary density can be used as an early metric for the mortality rate associated with cardiovascular diseases \cite{Nama2011-cz}. 
Measuring the capillary density is a crucial component of understanding fibrotic diseases \cite{Goligorsky2010-ga}. 
In a study on chronic kidney diseases, a reduction in the capillary density over time was determined to be useful as an early indicator to prompt timely interventions ~\cite{Edwards-Richards2014-fj}. 
Moreover, measurement of the velocity of red blood cells within capillaries can help assess fluid homeostasis and the transcapillary fluid flux \cite{Shore2000-mu}. 
Alterations in the velocity of red blood cells within capillaries influence the volume of nutrients delivered to surrounding cells and the duration of exchange through microvascular walls to surrounding tissues\cite{Michel1997-pd}. 
The flow velocity in cerebral capillaries may be increased owing to a body’s response to regulate a decrease in oxygen tension in neurons  \cite{Wei2016-sy}.

A common theme in microvascular analysis studies is the time-consuming and laborious nature of the tasks, requiring clinicians to select capillaries and manually determine blood flow velocity. 
Such tasks are ophthalmologically strenuous and susceptible to errors and observer variations across different datasets. On average, a trained researcher requires 20 minutes to analyse  a 20-s-long microvascular video \cite{Hilty2019-sv}.
Time-consuming analyses and considerable training requirements hinder the integration of microvascular microscopy into routine clinical practice \cite{ince2018second}.
Moreover, although the flow velocity and capillary density are assessed regularly, intra-capillary flow heterogeneity and capillary hematocrit are not routinely assessed owing to the difficulty in their manual analysis.
Thus, the aforementioned drawbacks also limit the number of parameters that can be analysed from microvascular videos \cite{ince2018second}.

The aim of this study is to relieve trained researchers of the burden of manual analysis, thereby enabling the trained researchers to focus on the development of methods and hypotheses.
The requirements of the system were determined based on two sources—a set of interviews conducted by the authors with medical doctors working at ODI Medical AS (MedTech company responsible for the e-health industrial application), with several years’ experience in microcirculation analysis, and the literature provided in three meetings organised by international experts in microvascular microscopy.
This series of meetings concluded that an automatic assessment of microcirculation is essential to integrate microvascular microscopy into clinical practice \cite{ince2018second}.

Thus, this paper presents a novel system for automatically assessing microcirculation by using microvascular videos. The primary requirements for a production-grade system are as follows:

\begin{enumerate}
\item reduction in the time required for microcirculation analysis from 20 minutes (manual analysis) to couple of seconds;
\item an accuracy of 93\% exceeding that of a trained researcher, which is 84\%;
\item detection and quantification of capillary density;
\item detection and quantification of capillary hematocrit;
\item tracking intra-capillary flow heterogeneity and direction of flow within capillaries;
\item classification of the velocity of red blood cell flow within the capillaries;
\item low power consumption, so that it can be used in battery-powered devices in hospitals (e.g., it should run on a CPU, rather than only on power-intensive  GPUs) .
\end{enumerate}

Relevant methods previously reported in the literature were either slow (requiring several minutes to analyse microcirculation videos) or exhibited poor accuracy that was unsuitable for clinical application. 
The proposed method is unique because it combines traditional methods with deep neural networks (DNNs) to obtain a relatively fast system with high accuracy.
The proposed method has already been applied and evaluated in various clinical studies (e.g., on COVID-19, pancreatitis, and acute heart diseases) \cite{ODIWebsite}.

The remainder of the paper is structured as follows.  
Section~\ref{related_work} presents the previous systems developed to analyse microcirculation images. 
Section~\ref{proposed_system} introduces the method of the proposed system, CapillaryNet.
Section~\ref{implementation} describes the system implementation and the data acquisition and annotation methods. 
Section \ref{resultsandDiscussion} provides the results obtained by using the proposed system and a discussion on the findings. 
Section~\ref{conclusion} concludes the paper.

\section{Related Work}
\label{related_work}

Dobble \etal{}~\cite{dobbe2008measurement} uses a frame averaging method to remove the 
plasma and white blood cells gaps within the capillary before using an algorithm to 
detect capillaries. Using frame averaging can lead to a lower overall density calculation 
 capillaries with most gaps or not enough blood flow will be disregarded. 
Furthermore, Dobble \etal{}~\cite{dobbe2008measurement} removes capillaries that are out 
of focus  which were considered to add noise to the frame averaging method. From our 
experiments with handheld microscopy, the nature of the rounded lens may lead to 40\% 
out-of-focus images on both edges of the video. It is very challenging to have a fully 
focused video the whole time, and some parts can always be out of focus. Therefore, this 
will further significantly reduce the capillary density values.

Hilty \etal{}~\cite{hilty2019microtools} has a very similar flow as
Dobble\etal{}~\cite{dobbe2008measurement} with minor changes.
Hilty \etal{}~\cite{hilty2019microtools} 
detects capillaries by first generating a mean image across all frames and then passing 
the resulting image to two pipelines, firstly classifying vessels of 20–30 $\mu$m in 
diameter as capillaries and secondly any vessel of up to 400$\mu$m in diameter as venules. 
The capillaries are then passed to a modified curvature-based region detection algorithm 
\cite{deng2007principal} to an image that has been stabilized and equalized with an adaptive
histogram. The result is a vessel map containing centerlines across structures 
between 20–30$\mu$m wide. As stated by the authors of the curvature-based region detection 
algorithm \cite{deng2007principal}, this type of detection is unintelligent and can 
lead to detecting artifacts such as hair or stains of similar sizes. Furthermore, 
due to the challenges in the skin profile stated above, the mean of the images across the 
whole video is not always the best representation value because different parts might have different lighting or capillaries that can be out of the optimal focus. 
Moreover, slight motion videos will have to be disregarded entirely because 
the central line is calculated across all frames instead of per frame. 

Similar to Dobble \etal{}~\cite{dobbe2008measurement}, Bezemer \etal{}~\cite{bezemer2011rapid} 
improves the method by using 2D cross-correlation to fill up the blood flow gaps caused by 
plasma and white blood cells. Using cross-correlation is a better method because the number of frames to be disregarded is reduced. However, 2D cross-correlation assumes a uniform blood flow and does not consider the dynamic change of flow between the gaps that can inherently decrease the accuracy of prediction.

Tam \etal{}~\cite{tam2010noninvasive} detect capillaries through a semi-automated method which requires the user to select points on the image. The algorithm then decides if there is a capillary present. Because this method relies on the user to select the capillaries, it cannot be used in a clinical environment due to the time of analysis of a microscopy video.

Demir \etal{}~\cite{demir2012automated} uses a Contrast Limited Adaptive Histogram Equalization method (CLAHE) \cite{reza2004realization} with a median filter and an 
adjustable threshold to detect capillaries on the weighted mean of five consecutive 
frames. These methods need to be adjusted accordingly depending on the 
illumination on the video and thickness of the skin. Such adjustments introduce a manual job where the user has to find the right combination of values for different videos 
or videos with different illumination. 

Cheng \etal{}~\cite{cheng2015reproducible} applies an image enhancement step followed by the manual highlighting by the user of the capillaries. The image enhancement process darkens the capillaries and increases the brightness of the background using a best-fit histogram method. Using their system, the user can further increase the contrast and smoothen the images manually to increase the differentiation of the capillaries from the background. These modifications can then be generated and applied to all future captured microscopy videos. However, this macro generation assumes that the videos will be captured with the same brightness and skin thickness. Moreover, the image used is in grayscale; therefore, if there are any artifacts, it can be mistaken for capillaries.

Tama \etal{}~\cite{tama2015nailfold} uses binarization followed by skeleton extraction and segmentation to quantify the capillaries. The binarization is applied to the green channel  it was assumed it has the highest contrast between the capillaries and the background. The Top-Hat transform method was used to reduce uneven illumination, followed by Wiener filtering to remove noisy pixels and then the Gaussian smoothing method to smoothen the image. The OTSU thresholding method is then applied to segment the capillaries from the background. This method relies on the user finding a reference frame from the video with the highest contrast.

Prentašic \etal{}~\cite{prentavsic2016segmentation} used a customized neural network to segment the foveal microvasculature. Their neural network consists of 3 Convolutional Neural Networks (CNN) blocks coupled with max-pooling and a dropout layer followed by two dense layers. Their neural network was trained in 30 hours, and the segmentation took approximately 2 minutes per single image with an accuracy of 83\%. The time taken and high-end hardware used to analyze a single image make it unsuitable for clinical use  users.

Nivedha \etal{}~\cite{nivedha2016classification} used the green channel of the image and used a non-linear Support Vector Machine~\cite{noble2006support} to classify the capillaries. This method involved a manual step where the user had to crop the region of interest to improve the histogram equalization. Nivedha \etal{}~performed different experiments comparing different denoising filters such as Gaussian, Wiener, Median, and adaptive media and concluded that the Gaussian filter is the most suitable for their data. Furthermore, the authors compared different segmentation methods, including OTSU, K Means, watershed, and concluded that the OTSU method was the most suitable for their data. The segmented images were then passed to an SVM, and the authors achieved an accuracy of 83.3\%.

Geyman \etal{}~\cite{geyman2017peripapillary} takes more of a manual approach by first 
using software to click away from the major blood vessels and then applying hardcoded 
calculations to detect the total number of capillaries based on the number of pixels 
in the region of interest. Such manual clicks are highly susceptible to observer variations across different datasets. 

Java \etal{}~\cite{javia2018machine} modifies the ResNet18~\cite{he2016deep} to quantify capillaries and uses the first 10-layers of the architecture. The main limitation of the ResNet architecture is that images have to be resized to 224x224. However, most capillary images are less than 100x100. Such images have to be scaled up, making this method inefficient and using more resources than needed. The authors reported accuracy of 89.45\% on their data; however, ResNet 18 ~\cite{he2016deep} has 11 million trainable parameters, and with such scaling up, training time can be up several hours, and prediction time can be up to several minutes. The sheer number of parameters can make this slow and inefficient use within a clinical setting. The training and test time was not reported in the paper.

Hilty \etal{}~\cite{hilty2019microtools} uses a similar method to Cheng \etal{}, where the authors first apply an image enhancement process followed by the highlighting of the vessels of interest to quantify the capillary. The authors use a combination of Contrast-limited adaptive histogram and a combination of the first and second-derivative Gaussian kernel convolutions~\cite{hilty2019microtools} to quantify capillaries.
This method is susceptible to detecting artifacts as capillaries in the images will not differentiate between capillaries and other objects in the image. This paper did not report the accuracy, training, and test time.

Ye \etal{}~\cite{ye2020vivo} utilized the concept of transfer learning and used the inception Single Shot Detector V2 (SSD-inception v2)~\cite{8026312} to build their neural network. The SSD-inception v2 has high accuracy with reduced computational complexity making it suitable for capillary detection~\cite{barba2020deep}. On the other hand, the authors used a spatiotemporal diagram analysis for the flow velocity calculation. This method requires white blood cells or plasma gaps to detect the velocity accurately. Therefore, capillaries that lacked such characteristics had to be disregarded, reducing the overall efficiency of velocity classification. Furthermore, the paper's authors stated that the spatiotemporal method could be cumbersome and time-consuming. The accuracy was not reported in this paper.

Hariyani \etal{}~\cite{hariyani2020capnet} used a U-net architecture combined with a dual attention module~\cite{hu2018squeeze,woo2018cbam}. The images have to be resized to 256x256, and an accuracy of 64\% was reported. This accuracy is too low for it to be used in a clinical setting.

Dai \etal{}~\cite{dai2020exploring} used a custom neural network similar to Prentašic \etal{} for segmentation. However, Dai \etal{} used five CNN blocks instead of three. Gamma correction and contrast limited adaptive histogram were used for equalization in the image enhancement step. The authors reported an accuracy of 60.94\%, which is too low to be used in a clinical setting.

Both Sherry G.Clendenon \etal{}~\cite{clendenon2019simple}, and Sang-Ho Park \etal{}~\cite{park2020} developed methods that can segment and quantify the microvascular structure using deep learning but had to use to inject the subject with ICG injection to enhance the capillaries so it can be detected easier with the algorithm.
The time for analysis was in minutes for both methods and involved an invasive procedure, meaning chemicals were added to the subject before taking a reading.
Anthony D Holley \etal{}~\cite{holley2021early}  used AVA (Version 4), which is based on Hilty \etal{}~\cite{hilty2019microtools,hilty2020automated} method described earlier. 

In the above, the more accurate methods require semi-automatic analysis while the more automatic methods are less accurate, making such methods unsuitable for a clinical setting. 
In contrast, CapillaryNet is fully automatic and able to detect capillaries in $\sim$0.9s\ with 93\% accuracy making it suitable to be used in a clinical setting.

\section{Methods}
\label{proposed_system}

\begin{figure}[!ht]
\center
\includegraphics[width=\linewidth]{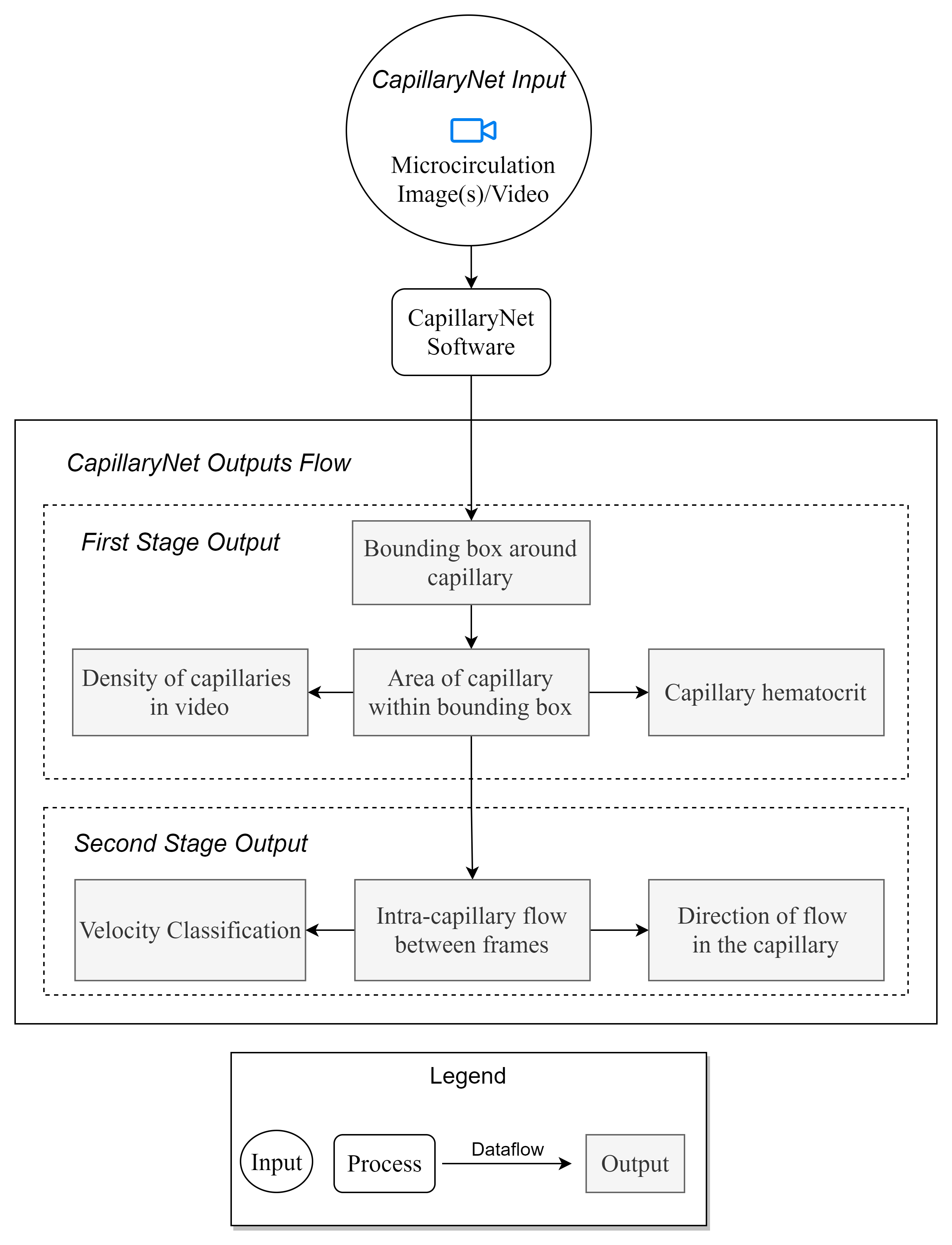}
\caption{Input and output of CapillaryNet. CapillaryNet accepts a video/image from the microscope as an input and outputs the depicted microcirculation parameters and values}
\label{CapillaryNetOverview_0}
\end{figure}

The system architecture of CapillaryNet is presented in this section. CapillaryNet comprises two primary functional stages. During the first stage, which is described in the first subsection, capillaries are detected, the area within the bounding box is calculated, and capillary hematocrit and capillary density are determined. During the second stage, which is described in the second subsection, the intra-capillary flow between frames is calculated, the direction of flow within the capillary is plotted, and the velocity within that capillary is calculated. The input and output of these two stages are shown in Figure \ref{CapillaryNetOverview_0}.

\subsection{First Stage - Capillary Detection, Area Quantification, and Density Calculation}

The first stage consists of three phases—generation of the regions of interest (RoIs), classification of the RoIs using CNN, and generation of masks on the classified RoIs. RoIs are defined as bounding boxes in specific areas of the image, suggesting the presence of a capillary. RoIs are detected via two independent pipelines: the hue, saturation, value (HSV)  pipeline and the Structural Similarity Index (SSIM) pipeline \cite{wang2004image}. Then, different enhancements are applied to the RoIs and these RoIs are transmitted to two different CNNs for capillary detection. Finally, the areas of RoIs containing capillaries are quantified during the mask generation phase and the capillary density of the image is calculated.

\subsubsection{HSV Pipeline of Phase 1: Generation of RoIs}

The first pipeline begins by applying the following three steps to smoothen the image. 
In the first step, the image is convolved using a 5 × 5 normalised box filter with a stride of one. 
The central pixel within that window is changed to the average value of the group within the window (see equation \ref{convolve}).
The image is then blurred using a 2D 9 × 9 Gaussian blur window (see equation \ref{gaussian}).
Gaussian filters are used because such filters do not overshoot the pixel value to a step function input while minimising the increase and decrease in pixel values.
Thus, in this case, sharp edges in the image are smoothened, while avoiding excessive blurring or the loss of edge information.
The image is then alpha-blended with the original input image using a Gaussian weight of -0.5 (see equation \ref{addweight}).
Alpha blending overlays a transparent version of the smoothened image on the original image as the background. 
It is an arithmetic operation that minimises noise while retaining the details of the image.
The transparency mask between these images is known as an alpha mask and a value of $\alpha=1.5$ is taken in this case. 
The aforementioned filters were selected owing to their relatively cheap computational requirements compared to a hybridised kernel \cite{srinivasu2021performance}.
Thus, using a simple Gaussian filter, the proposed method achieved the accuracy required for industrial and clinical application, while also satisfying duration and hardware requirements.
In comparison, hybridised kernels require greater computational capacity than that available in hospital environments.

\begin{equation}
K=\frac{1}{25}\begin{bmatrix}
1& 1 & 1 & 1 &1 \\ 
1& 1 & 1 & 1 &1 \\ 
1 & 1 & 1 & 1 &1 \\ 
1 & 1 & 1 & 1 & 1\\ 
1 & 1 & 1 & 1 & 1
\end{bmatrix} 
\label{convolve}
\end{equation}
where:
\begin{conditions}
 K     &  image smoothing factor
\end{conditions}

\begin{equation}
G(x,y)=\frac{1}{2\pi{\sigma ^{2}}}e ^\frac{x^{_{2}}+y^{_{2}}}{2\sigma ^{2}}
\label{gaussian}
\end{equation}
where:
\begin{conditions}
 x     &  distance from the origin in the horizontal axis \\
 y     &  distance from the origin in the vertical axis \\
 \sigma & standard deviation of the Gaussian distribution
\end{conditions}

\begin{equation}
\widehat{F}= (1-\alpha) F_{0}(x) + \alpha F_{1}(x), 
\label{addweight}
\end{equation}
where:
\begin{conditions}
 F_0     &  original image \\
 F_1     &  smoothing image
\end{conditions}

After the aforementioned steps are completed, the colour of the image is enhanced by a factor of three, and the contrast is enhanced by a factor of 2.5 \cite{clark2015pillow}.
The image is then converted to the HSV colour model, whose value represents the brightness of the colour on the spectrum from black to the average saturation value.
The hue channel represents the colours, e.g., red or yellow. The saturation channel represents the degree of colour on the spectrum from grey to the pure colour of the image.
In an image, the brightness, V, is calculated using Equation \ref{v} \cite{shaik2015comparative}:

\begin{equation}
V=max(R,G,B)
\label{v}
\end{equation}

The intermediate value, C, is then calculated by using Equation \ref{intermediate} \cite{shaik2015comparative}:

\begin{equation}
C=V-min(R,G,B)
\label{intermediate}
\end{equation}

C is used to derive the hue value and is calculated using Equation \ref{hue} \cite{shaik2015comparative}.

\begin{equation}
H=60^{\circ} \times \begin{cases}\text { undefined } & \text { if } C=0 \\ \left(\frac{G-B}{C}\right)(\bmod 6) & \text { if } V=R \\ \left(\frac{B-R}{C}+2\right)(\bmod 6) & \text { if } V=G \\ \left(\frac{R-G}{C}+4\right)(\bmod 6) & \text { if } V=B\end{cases}
\label{hue}
\end{equation}

Pixel saturation is calculated using Equation \ref{satur} \cite{shaik2015comparative}.

\begin{equation}
S= \begin{cases}0 & \text { if } V=0 \\ \frac{C}{V}, & \text { otherwise }\end{cases}
\label{satur}
\end{equation}

\begin{figure}[!ht]
\center
\begin{tabular}{cc}
\includegraphics[width=0.9\linewidth]{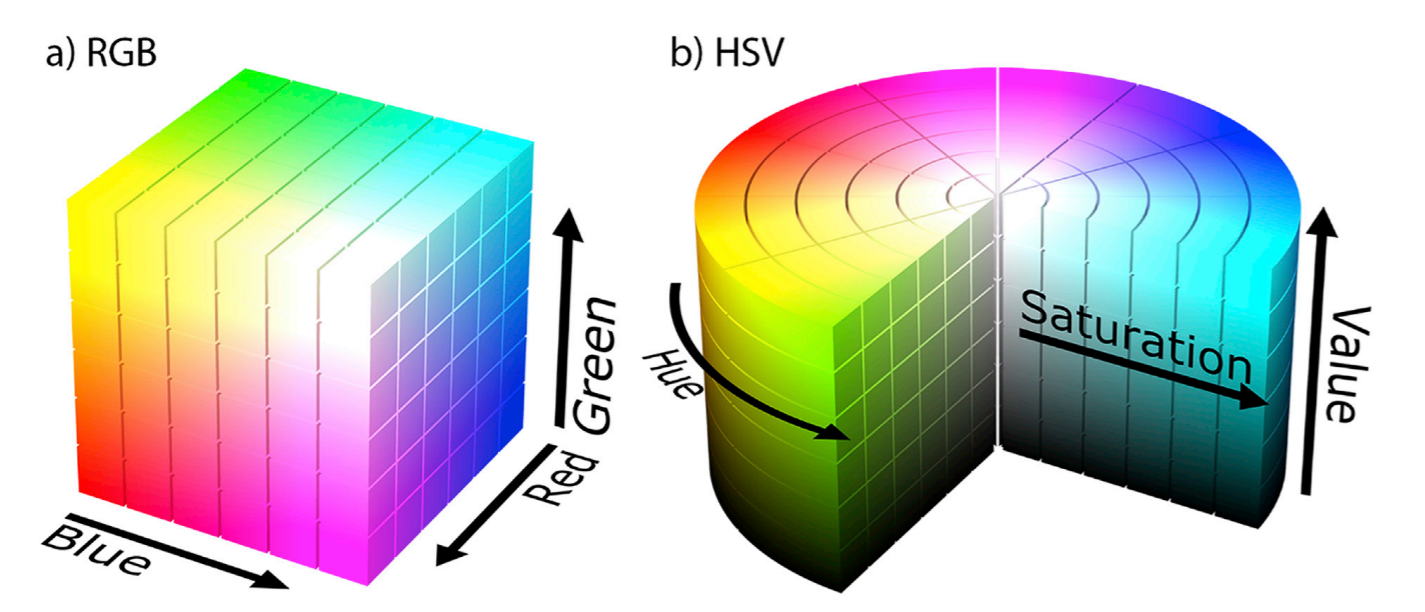}
\end{tabular}
\caption{Traditional RGB colour space is depicted on the left, whereas the HSV colour space is shown on the right. In RGB, the colour is determined by the resultant colour at a point within the space while in HSV the values of saturation, value and hue are combined to determine the colour. Images by Michael Horvath, available under Creative Commons Attribution-Share Alike 3.0 Unported license
}
\label{hsvRGB}
\end{figure}

The HSV colour model is used in this paper instead of the red, green, blue (RGB) model because of its greater robustness against changes in illumination and shadows \cite{shaik2015comparative}.
While the RGB model subtracts the value at each point to obtain the colour, the HSV model combines the values of hue, value, and saturation to obtain the colour value \cite{popov2018practices}. This fundamental difference in the determination of the colour of a pixel enables more accurate detection of capillary segments. The differences between HSV and traditional RGB colour spaces are depicted in Figure~\ref{hsvRGB}.

After several experiments, it was observed that the HSV colour values of capillaries were between (155, 60, 0) and (180, 255, 255). However, these values are modifiable in the code of the proposed methodology and can be modified by the user. The bitwise AND operator is used to calculate the conjunction of pixels between the specified HSV range and the image. This operation only considers standard pixels between the HSV range and the image. Subsequently, the remaining pixels are removed. The borders between the removed pixel and the HSV range are detected using a combination of the OTSU  threshold and the contour approximation method \cite{berezsky2018image,opencv123123}.

\subsubsection{SSIM Pipeline of Phase 1: Generation of RoIs}

\begin{figure}[!ht]
\begin{tabular}{c}
\includegraphics[width=0.9\linewidth]{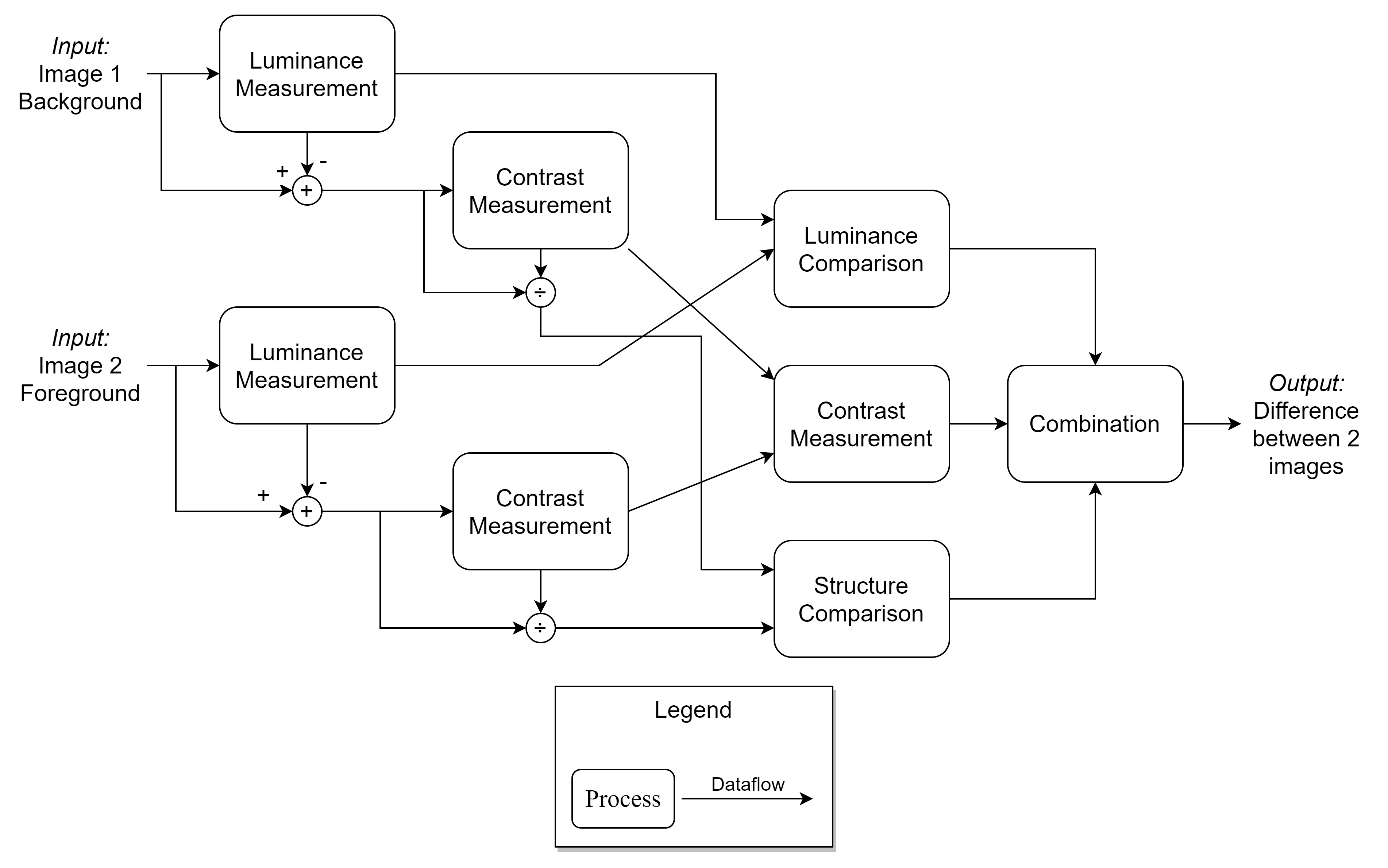}
\end{tabular}
\caption{Flowchart for the calculation of the Structural Similarity Index \cite{wang2004image}}
\label{SSIM_architecture}
\end{figure}

The second pipeline for RoI detection utilises the SSIM method \cite{wang2004image}, which extracts the background and subtracts it from the original image. Three sets of information are extracted from the image—luminance, contrast, and structure. The SSIM method is depicted in Figure 3 \ref{SSIM_architecture}~\cite{wang2004image}.

Luminance is measured by averaging the pixel in a given window. An 11 × 11 window is used in this case which is denoted by N in Equation \ref{luminance}.

\begin{equation}
\mu_{x}=\frac{1}{N} \sum_{i=1}^{N} x_{i}
\label{luminance}
\end{equation}

Contrast is measured by calculating the standard deviation of the pixel values (see Equation \ref{contrast}).

\begin{equation}
\sigma_{x}=\left(\frac{1}{N-1} \sum_{i=1}^{N}\left(x_{i}-\mu_{x}\right)^{2}\right)^{\frac{1}{2}}
\label{contrast}
\end{equation}

Structural comparison is performed by dividing the image in terms of its standard deviation along both input signals—first along the X-axis using the following equation:

\begin{equation} \left(\mathbf{x}-\mu_{x}\right) / \sigma_{x}\end{equation} 

and then along the Y-axis using the following equation:

\begin{equation} \left(\mathbf{y}-\mu_{y}\right) / \sigma_{y}\end{equation}

Each aforementioned step is completed by comparing the background image with the original image. The luminance comparison function shows the difference in brightness between two images (see Equation \ref{luminance_comparison}).

\begin{equation}
l(\mathbf{x}, \mathbf{y})=\frac{2 \mu_{x} \mu_{y}+C_{1}}{\mu_{x}^{2}+\mu_{y}^{2}+C_{1}}
\label{luminance_comparison}
\end{equation}

The C1 factor in the luminance comparison equation is calculated by multiplying the pixel value by a constant (see equation \ref{C_1}). The value of the constant $K_1$ is empirically determined based on the dataset.

\begin{equation}
C_{1}=\left(K_{1} L\right)^{2}
\label{C_1}
\end{equation}

The contrast comparison between the two images is calculated using standard deviation (see Equation \ref{contrast_comparison}).

\begin{equation}
c(\mathbf{x}, \mathbf{y})=\frac{2 \sigma_{x} \sigma_{y}+C_{2}}{\sigma_{x}^{2}+\sigma_{y}^{2}+C_{2}}
\label{contrast_comparison}
\end{equation}

The C2 factor in the contrast comparison equation is calculated by multiplying the pixel value by a constant (see Equation \ref{C_2})). The value of the constant $K_2$ is empirically determined based on the dataset.

\begin{equation}
C_{2}=\left(K_{2} L\right)^{2}
\label{C_2}
\end{equation}

The structure comparison function is also calculated using the standard deviations corresponding to the images (see Equation \ref{structure_compare}).

\begin{equation}
s(\mathbf{x}, \mathbf{y})=\frac{\sigma_{x y}+C_{3}}{\sigma_{x} \sigma_{y}+C_{3}}
\label{structure_compare}
\end{equation}
where:
\begin{eqnarray*}
&&\sigma_{x y} =  ~\mbox{see equation}~ \ref{sigma_equation} \\
&&C_{3} = C_2/2
\end{eqnarray*}

and

\begin{equation}
\sigma_{x y}=\frac{1}{N-1} \sum_{i=1}^{N}\left(x_{i}-\mu_{x}\right)\left(y_{i}-\mu_{y}\right)
\label{sigma_equation}
\end{equation}

\medskip

The SSIM is then deduced by calculating the difference in luminance, contrast, and structural difference (see equation \ref{SSIM_1}).
As deduced by the author of the SSIM paper~\cite{wang2004image}, assuming $\alpha$ = $\beta$ = $\gamma$ $=1$  and $C_3=C_2/2$, the equation can be simplified as shown in equation \ref{SSIM_2}.

\begin{equation}
\operatorname{SSIM}(\mathbf{x}, \mathbf{y})=[l(\mathbf{x}, \mathbf{y})]^{\alpha} \cdot[c(\mathbf{x}, \mathbf{y})]^{\beta} \cdot[s(\mathbf{x}, \mathbf{y})]^{\gamma}
\label{SSIM_1}
\end{equation}

\begin{equation}
\operatorname{SSIM}(\mathbf{x}, \mathbf{y})=\frac{\left(2 \mu_{x} \mu_{y}+C_{1}\right)\left(2 \sigma_{x y}+C_{2}\right)}{\left(\mu_{x}^{2}+\mu_{y}^{2}+C_{1}\right)\left(\sigma_{x}^{2}+\sigma_{y}^{2}+C_{2}\right)}
\label{SSIM_2}
\end{equation}

\medskip

Following the aforementioned steps, the mean structural similarity index between the resulting image and the original image is calculated, and all differences between these images are assigned bounding boxes and considered to be RoIs.
Each RoI is enhanced using pixel cumulative histogram equalization.
Combination of the two aforementioned pipelines increases the probability of capillary detection, thereby improving the overall accuracy of the method.
Moreover, the utilization of both pipelines is relatively inexpensive compared to purely CNN-based detectors.

\begin{figure}[!ht]
\center
\begin{tabular}{cc}
\includegraphics[width=0.9\linewidth]{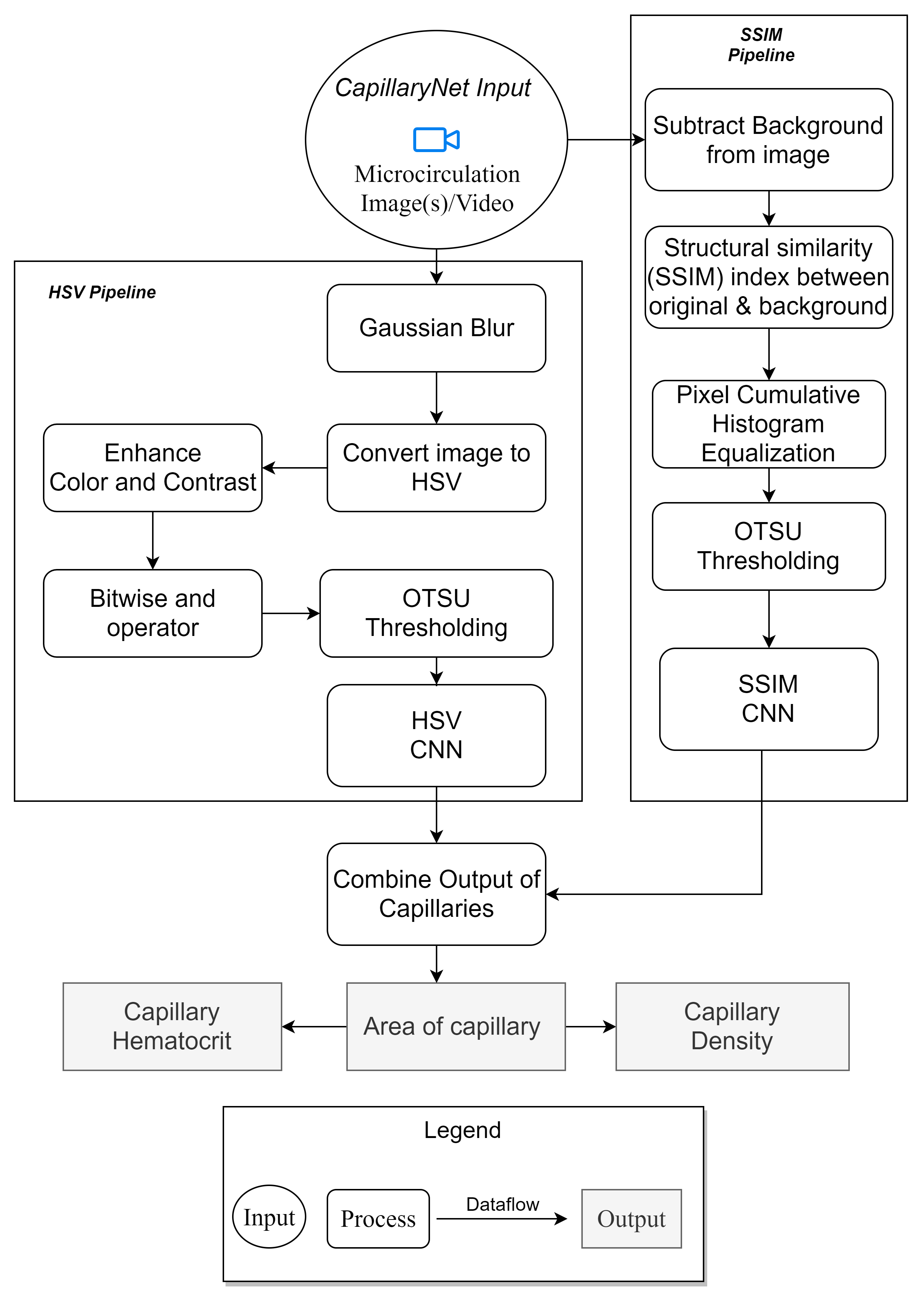}
\end{tabular}
\caption{
Overall flowchart for the generation of RoIs in phase 1
}
\label{RoIGeneration_overall}
\end{figure}

The overall flowchart for phase 1 (HSV and SSIM), as described in this subsection and the previous ones, is presented in Figure \ref{RoIGeneration_overall}.

\begin{figure*}
\resizebox{\textwidth}{!}{%
\begin{tabular}{ccc}
\multicolumn{3}{c}{
\adjincludegraphics[height=2cm,width=3cm,trim={0 0 0 0},clip]{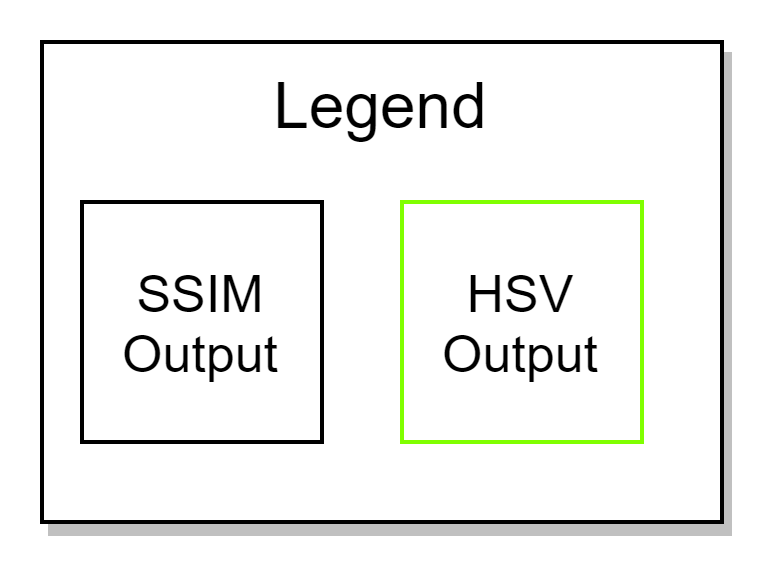}
}
\\
\adjincludegraphics[height=4cm,width=6cm,trim={0 0 0 0},clip]{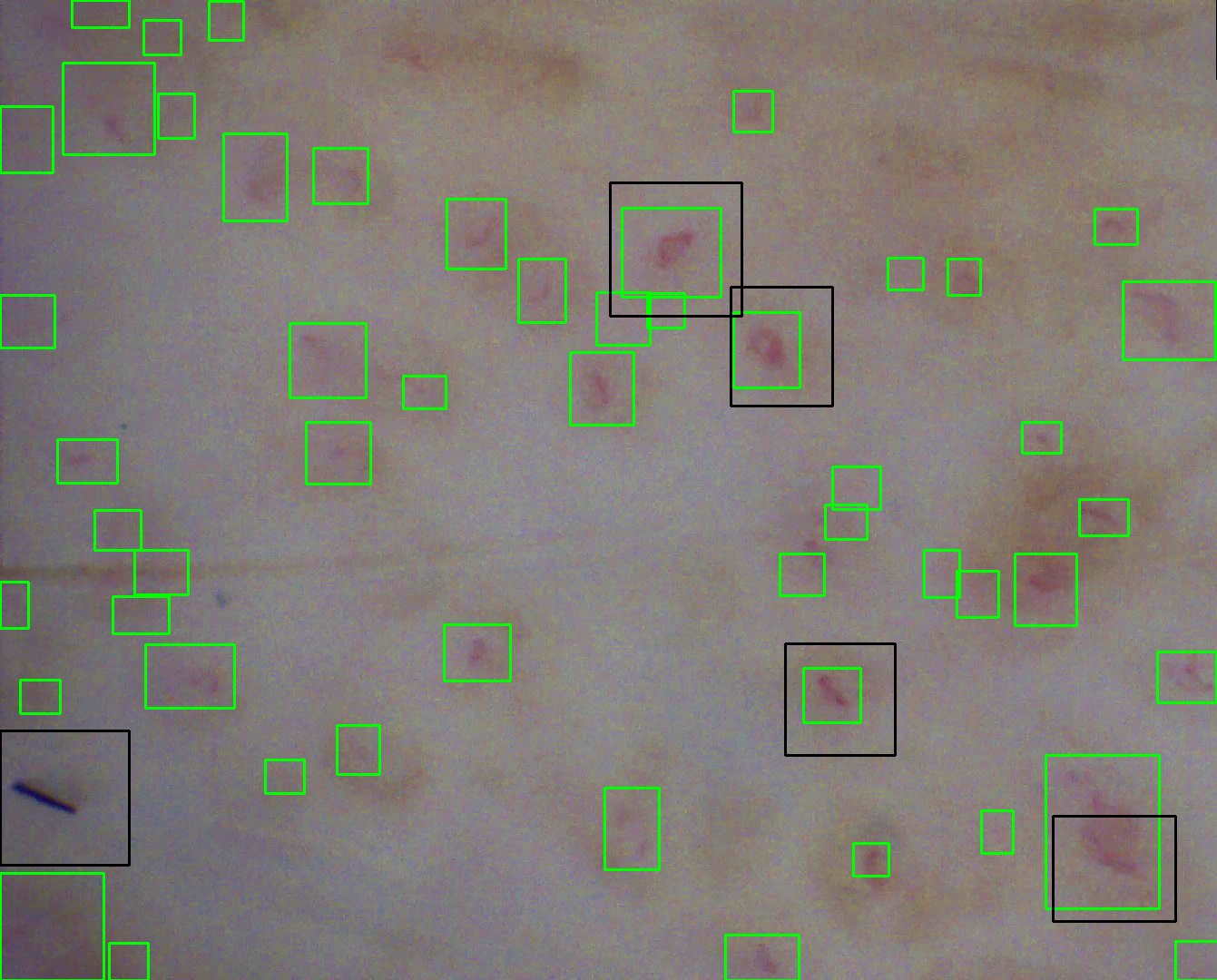}
&
\adjincludegraphics[height=4cm,width=6cm,trim={0 0 0 0},clip]{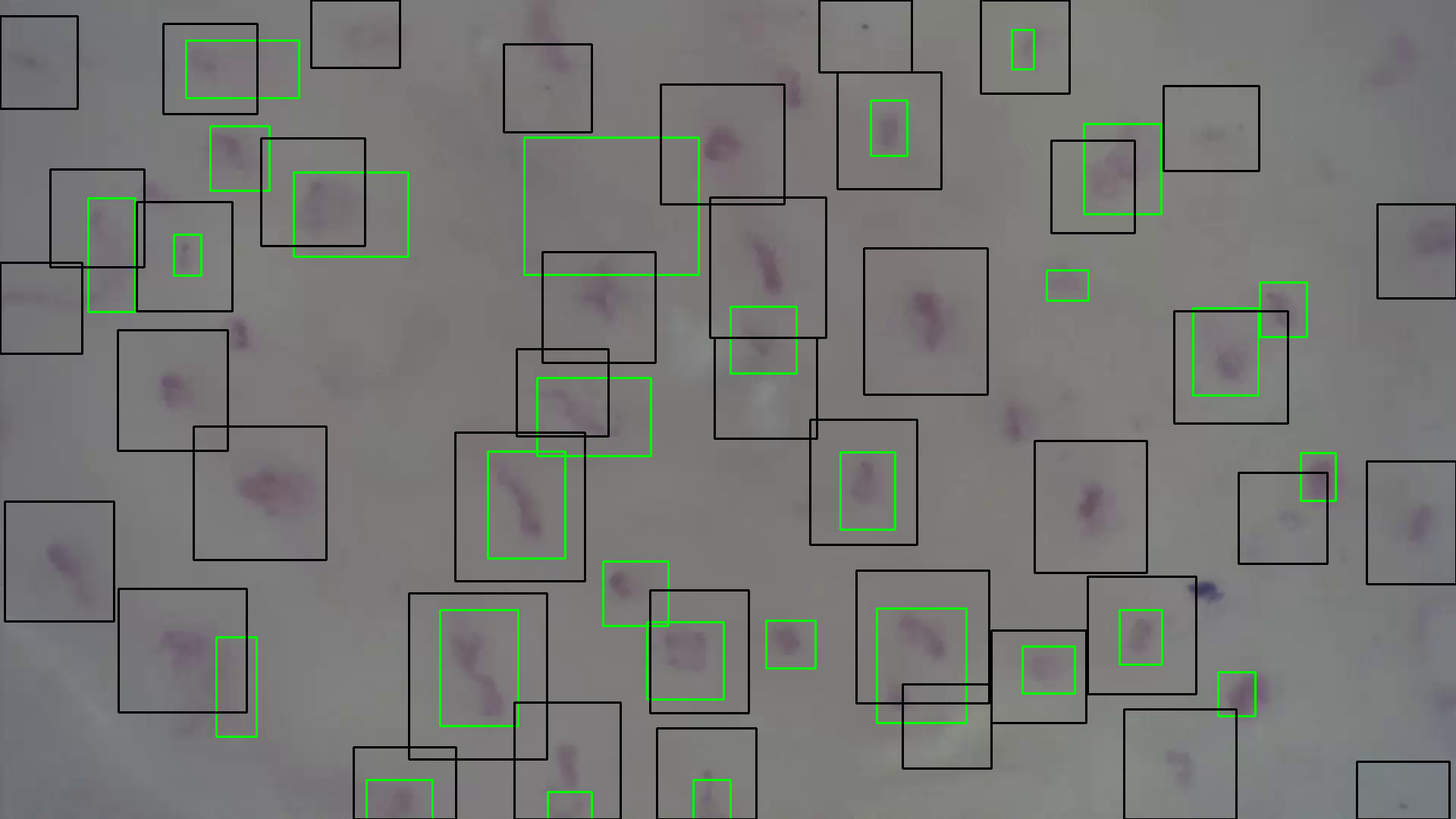}
&
\adjincludegraphics[height=4cm,width=6cm,trim={0 0 0 0},clip]{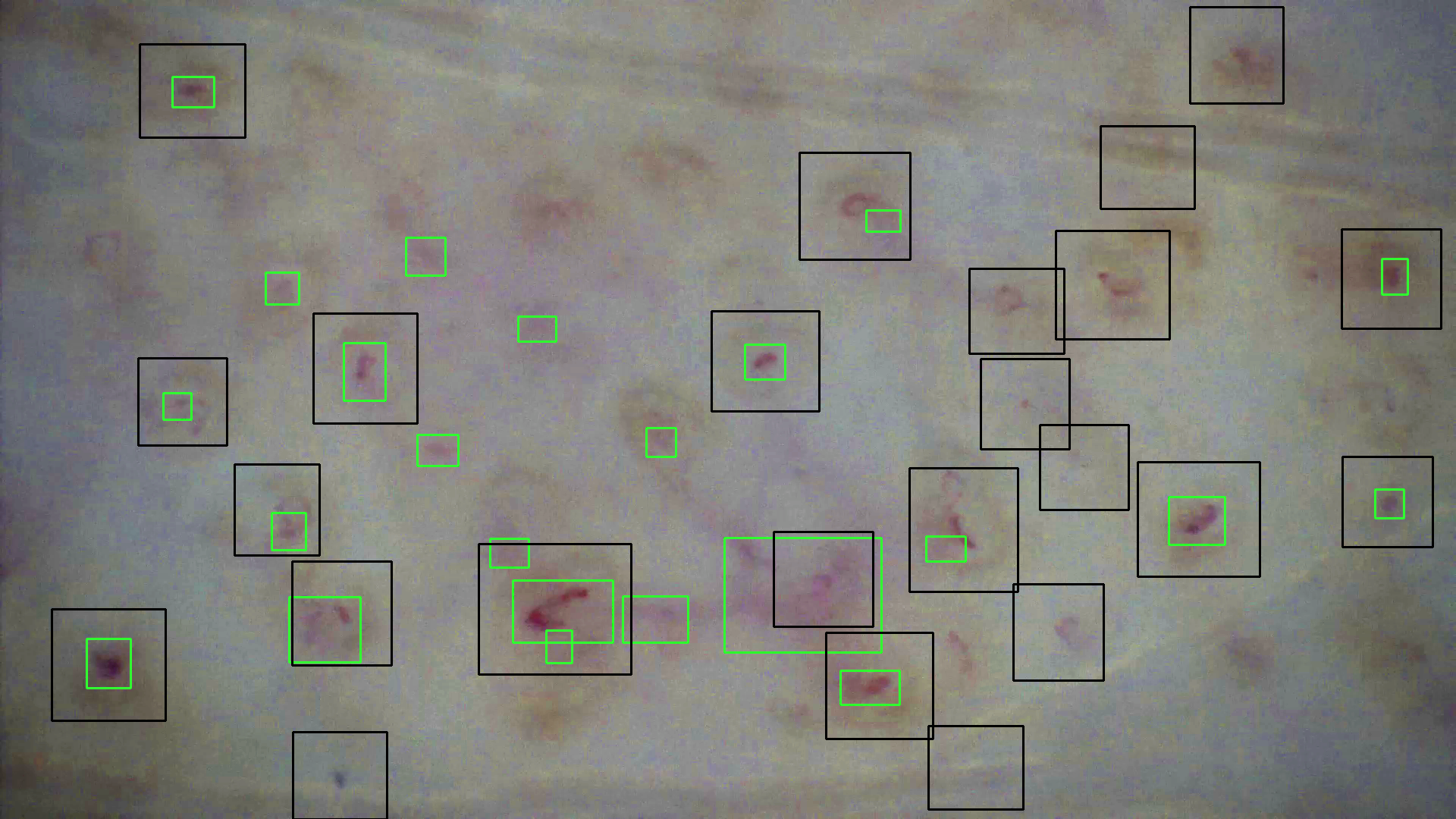}
\\
a     & b     & c   
\end{tabular}%
}
\caption{
(a) The HSV pipeline (green boxes) detects more RoIs than the SSIM pipeline (black boxes). 
(b) The SSIM pipeline detects more RoIs than the HSV pipeline. 
(c) On the right side of the image, SSIM detects more RoIs than HSV, and on the left side of the image, HSV detects more RoIs than SSIM.
Thus by combining both pipelines, the probability of detecting capillaries is increased, thereby improving the overall accuracy of the method
}
\label{output_example_hsv_ssim}
\end{figure*}

In Figure \ref{output_example_hsv_ssim}, the RoIs detected by using HSV are highlighted in green, while those detected using SSIM are highlighted in black. 
Figure \ref{output_example_hsv_ssim}a
shows a case where a greater number of RoIs is captured using the HSV pipeline than SSIM, Figure \ref{output_example_hsv_ssim}b depicts a case in which a greater number of RoIs is detected using SSIM compared to HSV, and Figure \ref{output_example_hsv_ssim}c reveals that SSIM and HSV pipelines capture RoIs in different parts of the image. 

Every bounding box in the image is considered to be a RoI. Each RoI is transmitted to the CNN (in phase 2) for classification — this process is described in the next section.

\subsubsection{Phase 2: CNN}

The second stage begins with the transmission of the RoIs from the SSIM and HSV pipelines to the CNN of CapillaryNet for capillary detection.
Each pipeline is assigned its own CNN model.

\begin{figure}[!ht]
\begin{tabular}{cc}
\includegraphics[width=0.9\columnwidth]{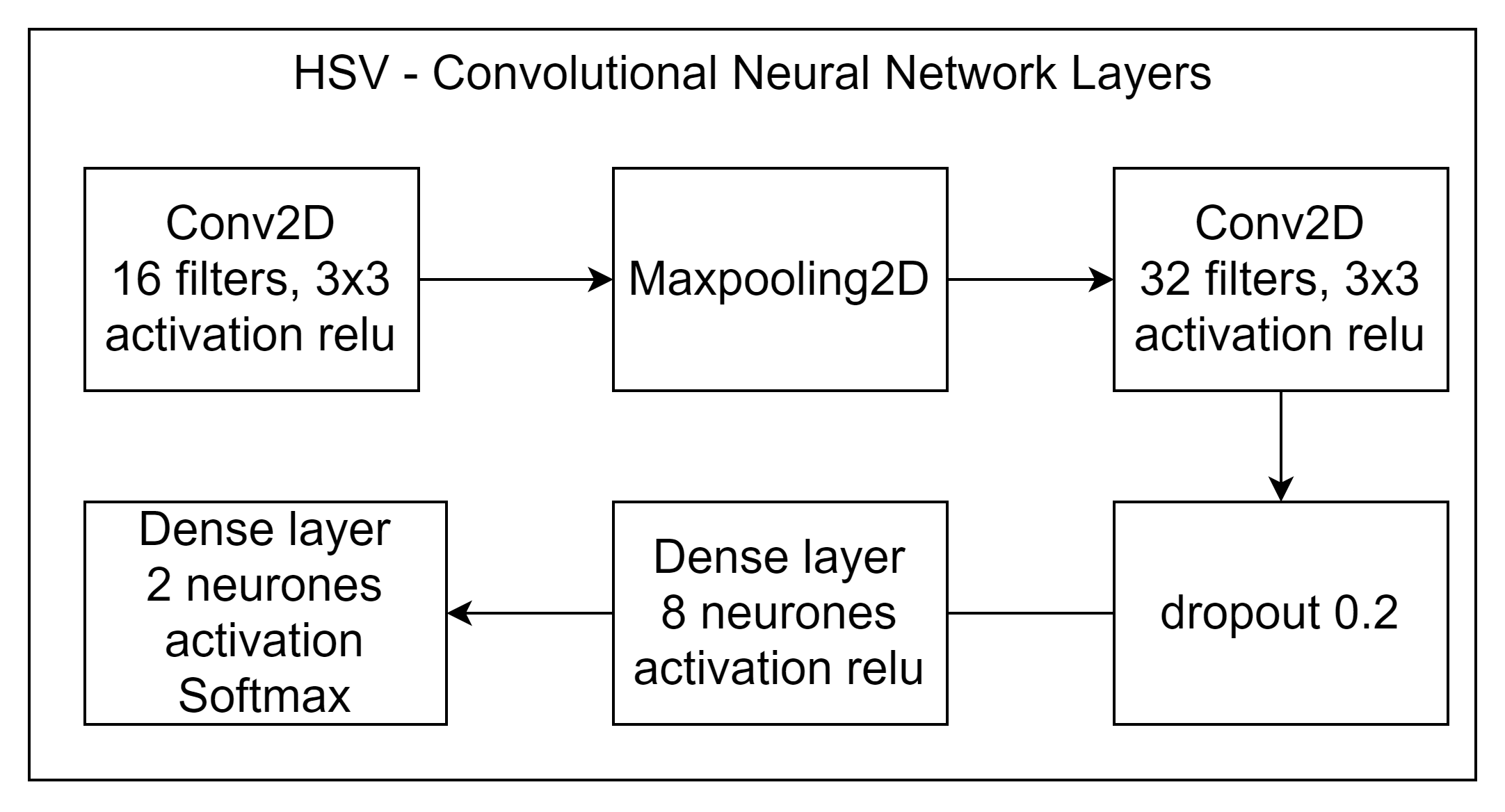}
\end{tabular}
\caption{CNN architecture corresponding to the HSV colour space pipeline with stride equal to 1 for the convolutional layers}
\label{CNN_Model_HSV}
\end{figure}

The model corresponding to the HSV pipeline comprises 7,418 parameters.
The minimum size of input image to ensure high detection accuracy for this CNN was experimentally deduced to be 15 × 15.
The HSV CNN architecture includes  two CNNs with 16 and 32 filters, respectively.
This is followed by a dropout layer of 0.2 and 2 dense layers with 8 and 2 neurons, respectively.
Both the convolutional layers and the first dense layer contain a ReLU activation function ~\cite{agarap2018deep} while the last layer contains a Softmax activation function (see Equation ~\ref{softmax}).
The Softmax function scales the sum of the values to one, enabling the interpretation of the outcomes as probabilities when the value exceeds 0.5, the corresponding RoI is predicted to contain a capillary, while a value less than 0.5 indicates a high probability that the corresponding RoI does not contain a capillary.
The CNN architecture corresponding to the HSV pipeline is displayed in Figure \ref{CNN_Model_HSV}.

\begin{equation}
\sigma\left(x_{j}\right)=\frac{e^{x_{j}}}{\sum_{i=1}^2 e^{x_{i}}}
\label{softmax}
\end{equation}

The CNN is optimised by using sparse categorical cross-entropy and trained for 20 epochs.
Cross-entropy represents the loss between a label and its corresponding prediction.
Sparse categorical cross-entropy is a function in Tensorflow and refers to the labelling of the input dataset .
It is the same as cross-entropy loss, but takes only labels which are integers and not 1-hot vectors.
It is calculated on every iteration to minimise the loss, thereby increasing the overall accuracy of the neural network, and it is expressed by Equation ~\ref{sparse}. 
The classes in the dataset considered in this study are mutually exclusive—capillaries and non-capillaries.
The capillary category is assigned the label, 1, and the non-capillary category is labelled 0.

\begin{equation}
H(p, q)=-\sum_{x \in \text { classes }} p(x) \log q(x)
\label{sparse}
\end{equation}
where:
\begin{conditions}
p(x)     &  true probability \\
q(x)     &  model predicted probability
\end{conditions}

\begin{figure}[!ht]
\begin{tabular}{cc}
\includegraphics[width=0.9\columnwidth]{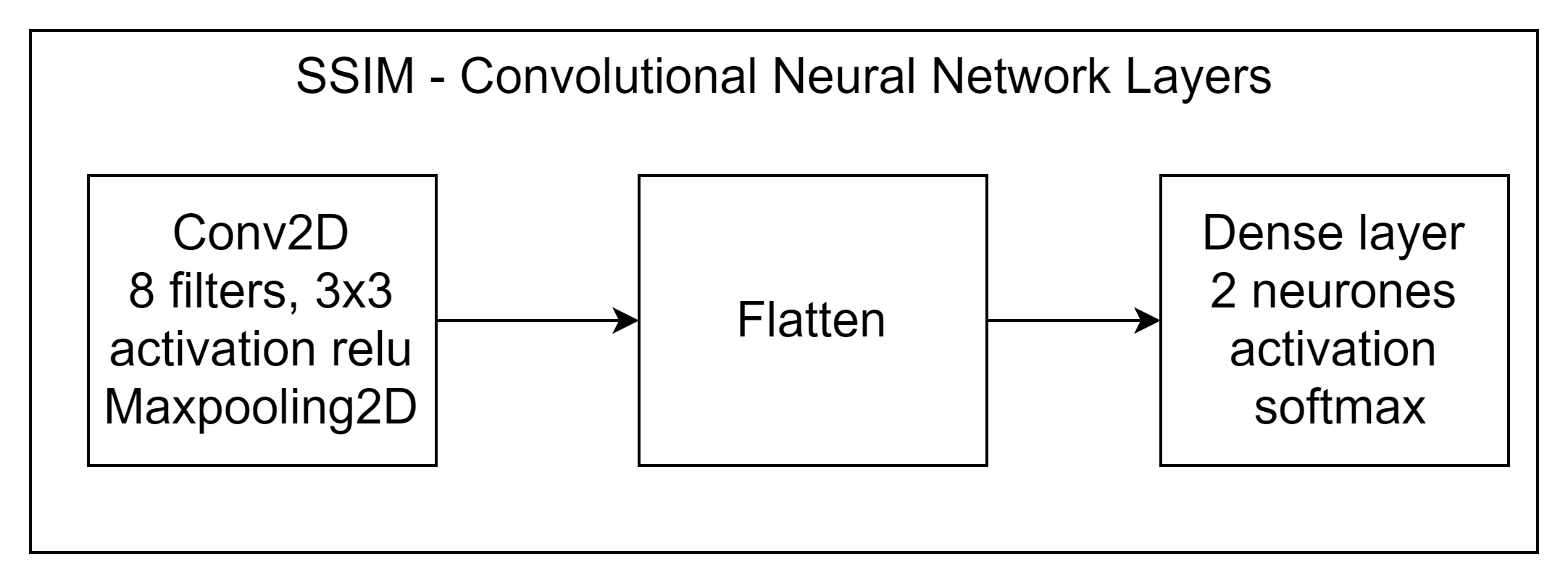}
\end{tabular}
\caption{CNN architecture of the SSIM Pipeline with stride equal 1 for the convolutional layer}
\label{capillaryNet_SSIM}
\end{figure}

The model architecture corresponding to the second SSIM pipeline includes one CNN with a ReLU activation function with eight filters ~\cite{agarap2018deep}, followed by one dense layer containing two neurons with a softmax activation function, as illustrated in Figure \ref{capillaryNet_SSIM}.
The input image is taken to be of size $50\times 50$, and 9,442 parameters are considered in aggregation.
The CNN is optimised using Adam \cite{kingma2014adam} and trained for 30 epochs. The loss of the model is computed in terms of cross-entropy loss between the labels and respective predictions. Table \ref{models_information} summarises the CNN parameters for both of these models.

The outputs of the two CNNs are then combined by integrating the coordinates of the capillaries into a single list.
This list contains the coordinates of all RoIs that are predicted to contain a capillary and is applied to the input image.

\begin{table}[!ht]
\caption{Parameters for the HSV CNN model and the SSIM CNN Model}
\label{models_information}
\resizebox{\columnwidth}{!}{%
\begin{tabular}{|c|c|c|}
\hline
\textbf{Parameter Name}   & \textbf{HSV CNN Model}          & \textbf{SSIM CNN Model}         \\ \hline
\textbf{Pad}              & half padded                     & not padded                      \\ \hline
\textbf{Image Dimension}  & 15 by 15                        & 50 by 50                        \\ \hline
\textbf{Layers}           & 3                               & 3                               \\ \hline
\textbf{CNN Filters}      & 16 and 32                       & 32 and 64                       \\ \hline
\textbf{Dense Layers}      & 8 with no dropout               & 64 with dropout 0.5             \\ \hline
\textbf{Activation Class} & Relu and Softmax                & Relu and Software               \\ \hline
\textbf{Optimizer}        & Adam                            & Adam                            \\ \hline
\textbf{Learning Rate}    & 0.001 to 1e-10         & 0.001 to 1e-12           \\ \hline
\textbf{Beta 1 Value}     & 0.9                             & 0.9                            \\ \hline
\textbf{Beta 2 Value}     & 0.999                           & 0.999                          \\ \hline
\textbf{epsilon value}    & 1e-07                           & 1e-07                           \\ \hline
\textbf{Loss Type}             & Sparse Categorical & Sparse Categorical \\ \hline
\textbf{Training Loss first Epoch}             & 0.3416 & 1.1332 \\ \hline
\textbf{Training Loss Last Epoch}             & 0.0783 & 0.1605 \\ \hline
\textbf{Validation Loss first Epoch}             & 0.2170 & 0.6783 \\ \hline
\textbf{Validation Loss Last Epoch}             & 0.0781 & 0.1729 \\ \hline

\end{tabular}%
 }
\end{table}

\subsubsection{Phase 3: Mask Generation}

\begin{figure}[!ht]
\begin{tabular}{l}
\includegraphics[width=0.9\columnwidth]{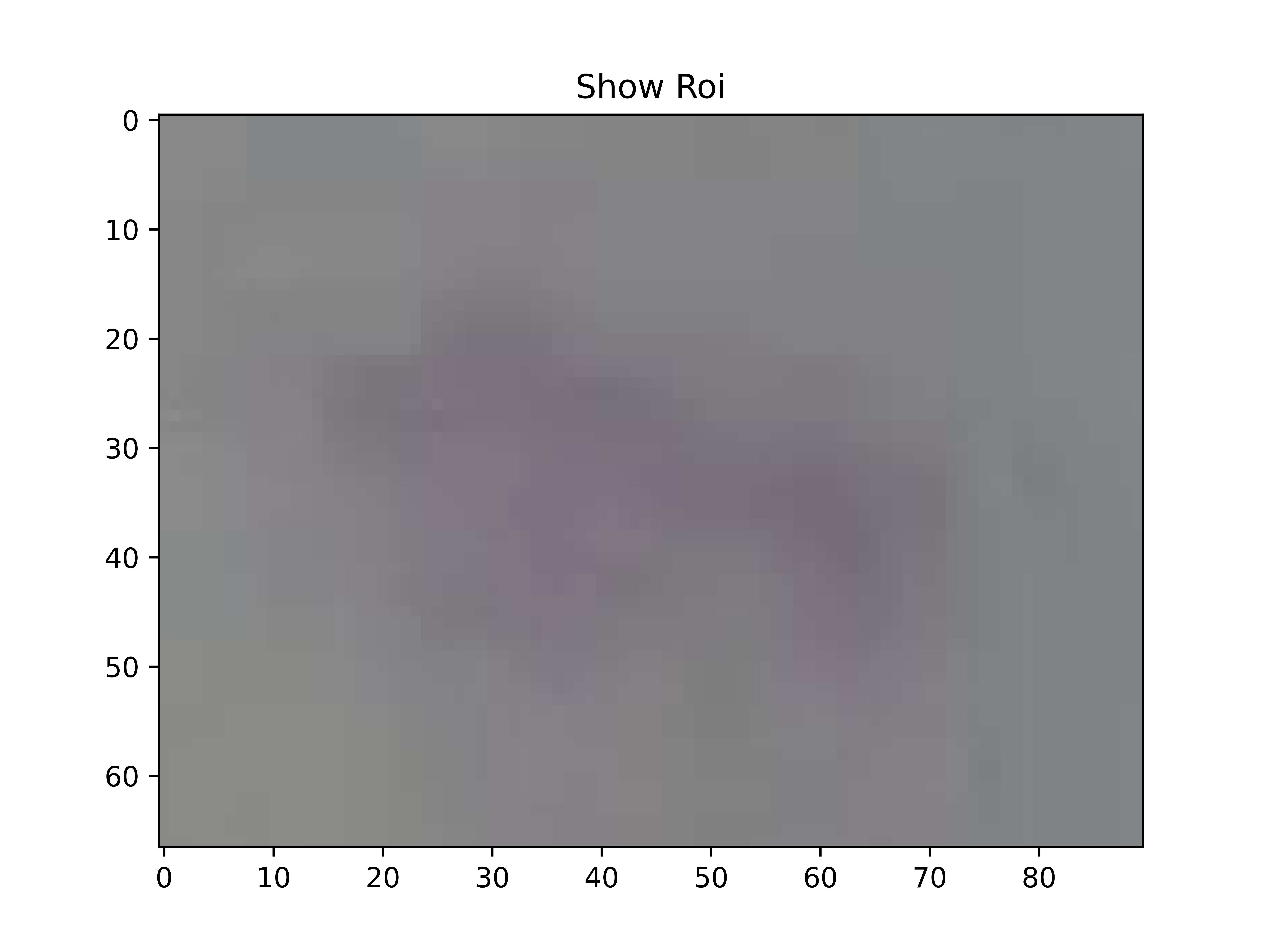}\\
\multicolumn{1}{c}{a} \\
\includegraphics[width=0.9\columnwidth]{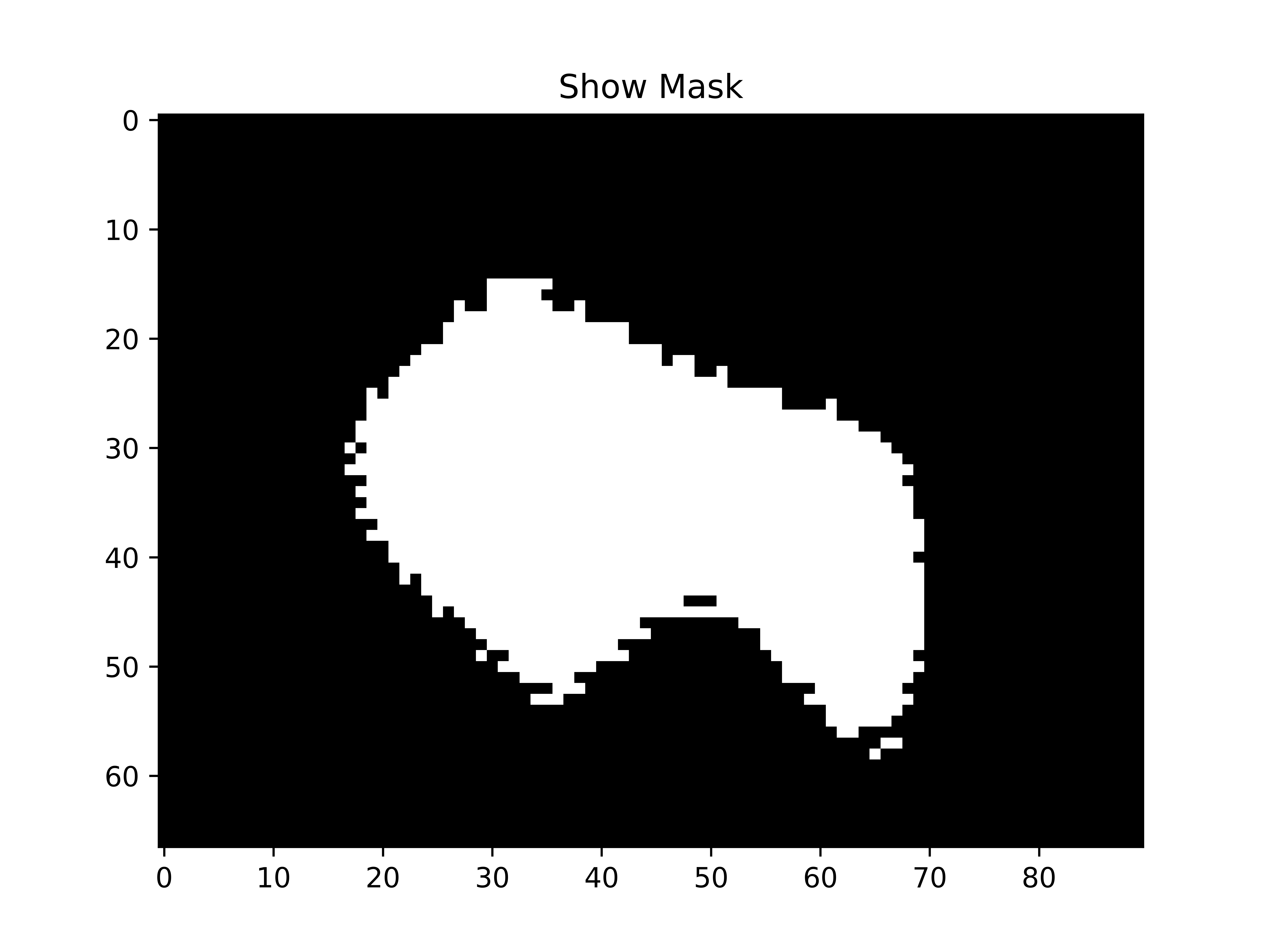}\\
\multicolumn{1}{c}{b}
\end{tabular}
\caption{(a) Original image of a RoI detected by deep learning architecture of CapillaryNet. (b) Area occupied by the capillary}
\label{rbc_images}
\end{figure}

Detected capillaries are processed using a Gaussian blur function, which reduces the pixelation of the image.
Gaussian blur is performed using a window of size 31 × 31.
Then, the red pixels are extracted from that image by applying an adaptive Gaussian threshold function that converts all red pixels to white ones and all other pixels to black ones.
The total amount of white pixels and the output corresponding to RoI including a single capillary are depicted in Figure \ref{rbc_images}a and Figure \ref{rbc_images}b, respectively.
The mask is then re-calculated for every consecutive frame of the video to determine capillary density at all time instances and the capillary hematocrit.
The re-calculation of the mask over the entire duration of the video enables heterogeneity of capillary perfusion, which allows the monitoring of red blood cell flows within capillaries.
Via the aforementioned method, CapillaryNet detects a capillary, the area it occupies, and the blood flow through it over time.
These pieces of information help clinicians assess the average capillary hematocrit and its variation over time.

\subsubsection{Final Remarks on the First Stage}

The detection of capillaries in microcirculation images using CapillaryNet, the calculation of the areas within the bounding boxes, and the calculation of the capillary hematocrit and capillary density as output of the first stage have been described in previous subsections.
The CapillaryNet architecture achieves these goals by uniquely combining traditional and deep learning models and tailoring the parameters for capillary detection.

In previous studies, as discussed in Section 2, either traditional methods or deep neural networks had been utilised. Although traditional methods are faster than deep neural networks, traditional methods exhibit poorer accuracy as deep neural networks can learn from trained data—purely traditional methods output all non-background regions (e.g., in our case, the skin) as capillaries. On the other hand, deep neural networks are resource-intensive and time-consuming. Therefore by assigning a significant portion of the task to traditional computer vision techniques and using deep neural networks solely for detection, CapillaryNet outperforms the methods discussed in Section 2.

In the next subsection, the calculation of intra-capillary flow between frames using CapillaryNet, plotting the flow direction within the capillary, and the classification of the velocity within the capillary are discussed.

\subsection{Second Stage - Velocity Classification and Determination of Flow Direction}

The capillaries detected by the CNN are transmitted to the second stage of CapillaryNet. During this stage, the architecture determines the velocity and flow direction of blood within the capillaries. First, certain enhancements are applied to the RoI region from the original frame. The velocity detection algorithm requires a different set of enhancements compared to the capillary detection algorithm. Then, the image is denoised and the RoI is transmitted to the velocity detection algorithm. These steps are discussed in further detail in the following subsections.

\subsubsection{Phase 1: Image Enhancement using Pixel Cumulative Histogram Equalization}

First, a median blur with a kernel size of 5 × 5 is independently applied to each channel. In each channel, the central pixel value within the kernel window is re-computed based on the median value of the pixels within the kernel window. The pixels in the image range between 0 and 255  and a histogram is obtained based on the modified RoIs, which illustrates the intensity distribution of the pixels. The cumulative distribution of the pixels is then calculated based on the histogram. Finally, a cut-off threshold is applied, and all pixels with values within 10\% of the cumulative increased distribution are removed. The threshold adjustment increases the overall brightness and contrast of the capillaries within the RoI, making the RoIs clearer and sharper. Adoption of this threshold is, therefore, critical for tracking the movement of pixels across different frames.

\subsubsection{Phase 2: Image Denoising using Non-local Means Denoising Algorithm}

The increase in the contrast and overall sharpness of each RoI is accompanied by an increase in the noise of each pixel.
In this step, the noise corresponding to each pixel is reduced while retaining all information in the pixel.
First, a median blur is applied to denoise each pixel locally, thereby increasing the overall smoothness of the image.
As the noise value varies across pixels, some areas of an image may require more extensive denoising than others.
After the computation of the noise variable as a non-local mean, the Gaussian distribution is applied and local noise is minimised.
This computation is performed by converting the image to the LAB colour space - where 'L' is the lightness, 'A' (green/magenta ) and 'B' (blue/yellow ) are the chromatic axes ~\cite{kaur2012comparison}.
The pixels are then then separately denoised on the L and AB components, thereby ensuring that each group of pixels in the image is treated in accordance with their local noise values rather than using the global noise value for the entire image.
This completes the enhancement and smoothening of RoIs.
In the next step, each RoI is processed using the velocity detection algorithm of CapillaryNet, and the velocity, infra-flow direction, and flow direction within the capillaries are classified.

\subsubsection{Phase 3: Velocity Detection}

\begin{figure}[!ht]
\begin{tabular}{c}
\includegraphics[width=0.9\columnwidth]{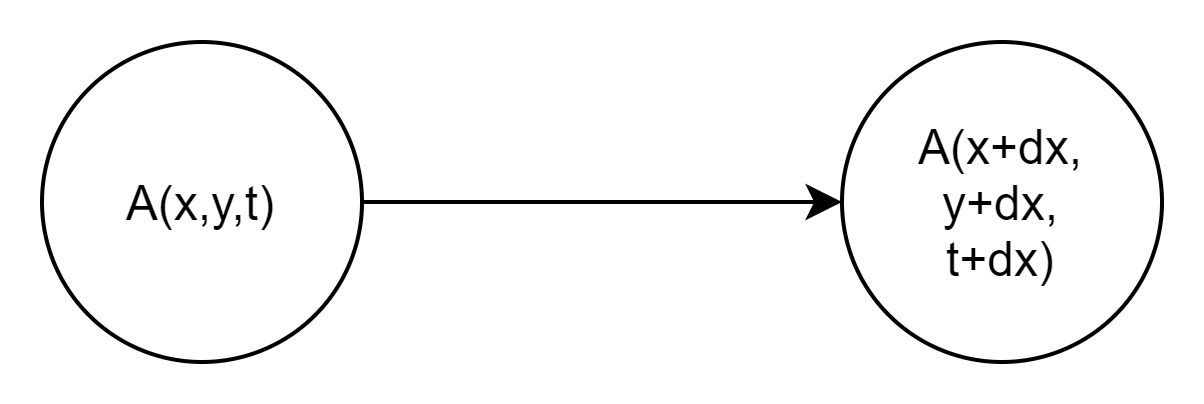}
\end{tabular}
\caption{Calculation of pixel displacement. A pixel at position (x,y), at time t, is displaced. The difference between the new and old positions, over time duration dt, is given by (dx, dy)}
\label{pixel_movement}
\end{figure}

The Gunnar Farneback \cite{farneback2003two} algorithm is a two-frame motion estimation algorithm, which performs approximation by calculating the quadratic polynomials and polynomial expansion transforms between each pair of successive frames.
The polynomial expansion coefficient is used to derive the displacement fields of each pixel, assuming pixel intensities to be constant between successive frames.
The velocity vectors are obtained based on the differences in the locations of the pixels between successive images.
A velocity vector represents the movement of an object from (x,y) in the direction, (dx,dy), in time, dt.
This movement is illustrated in Figure \ref{pixel_movement}.

The velocity vector is used to calculate the flow direction and the intra-capillary flow between frames, and perform velocity classification.
The algorithm assumes that the movement of an object can be described using a quadratic function $I(x,y,t)$ (given by Equation \ref{pixel_dis}).
The quadratic function is then expanded using a hierarchical scheme of separable convolutions utilizing Taylor series approximations of the right-hand side~\cite{corliss1982solving}, thereby yielding a system of linear equations in variables dx, dy, and dt.

The algorithm scales the images by a factor of 0.9 using a window of size, 10 × 10 with pyramid level 2. 
This calculation is performed over ten iterations at each pyramid level to increase the accuracy of the calculation.
A polynomial size of 5 and a polynomial sigma of 1.2 are considered.
The Cartesian coordinates are then converted to polar coordinates, which produces the magnitude and the angle of each 2D vector—here, the magnitude represents the absolute value of the velocity vector and the angle represents the direction of the flow.

\begin{figure}[!ht]
\center
\begin{tabular}{cc}
\includegraphics[width=0.9\columnwidth]{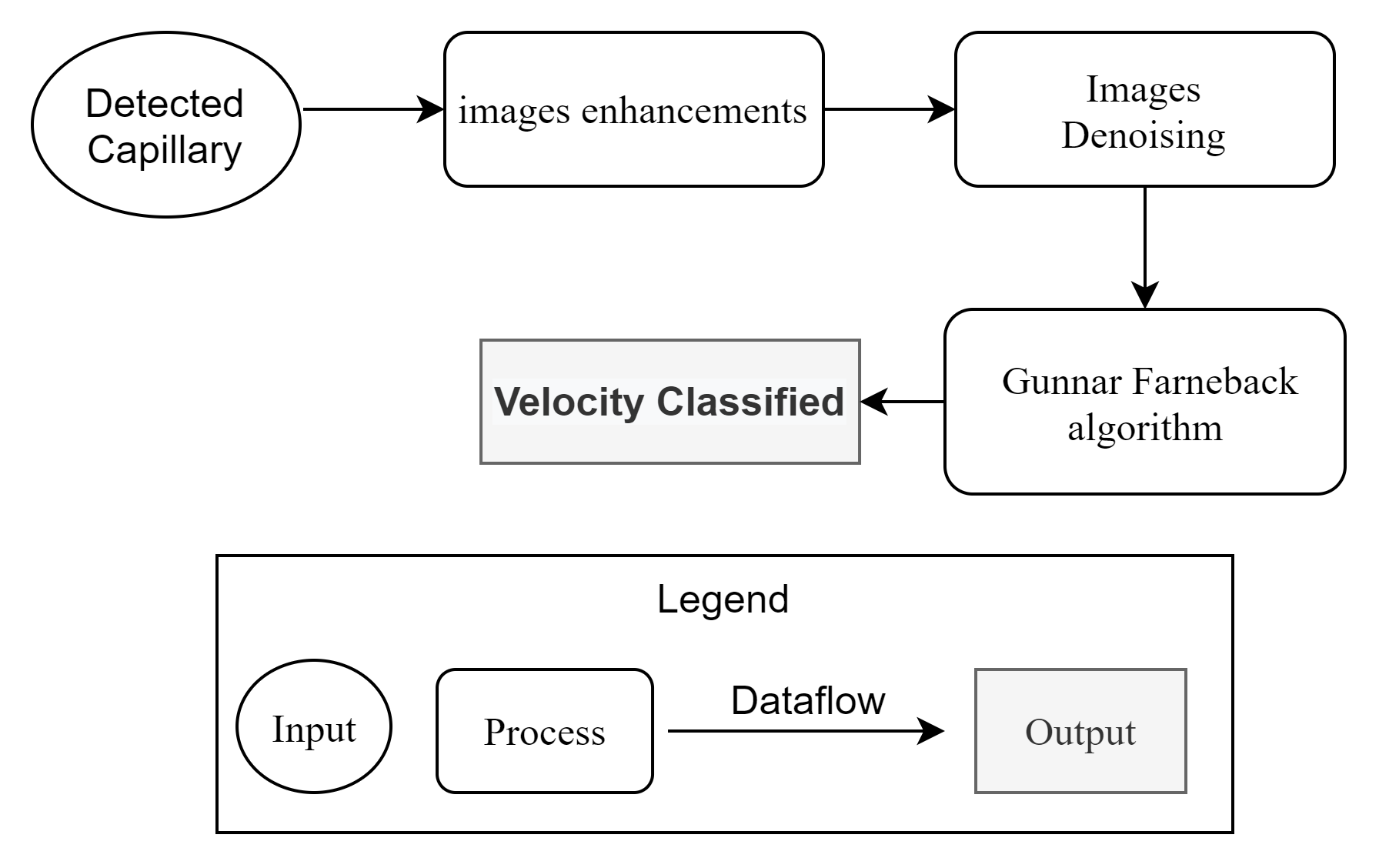}
\end{tabular}
\caption{CapillaryNet Architecture - Velocity Detection Flow}
\label{VV_Arch}
\end{figure}

The velocity vector represents the intra-capillary heterogeneity of the flow velocity value and the angle in the flow direction.
The averaged intra-capillary heterogeneity of the flow velocity vector yields the velocity classification of the RoI. The velocity detection method is illustrated in Figure \ref{VV_Arch}.

\begin{equation}
I(x, y, t)=I(x+d x, y+d y, t+d t)
\label{pixel_dis}
\end{equation}

\begin{figure}[!ht]
\center
\begin{tabular}{cc}
\includegraphics[width=0.5\columnwidth]{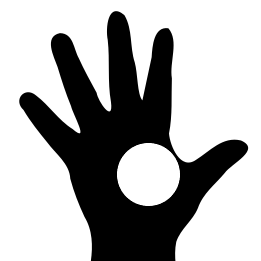}
\end{tabular}
\caption{Skin region where the microvascular videos were recorded is marked in white}
\label{dorsum_area_sampleA}
\end{figure}

\begin{figure}[!ht]
\center
\begin{tabular}{cc}
\includegraphics[width=0.9\columnwidth]{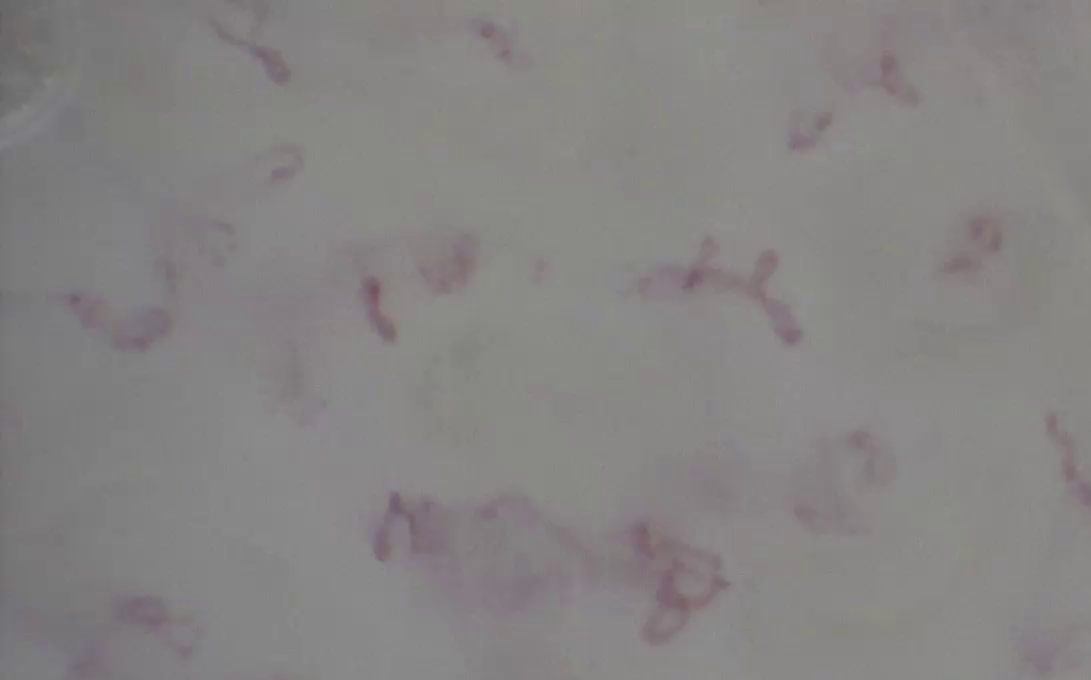}
\end{tabular}
\caption{Image from the captured video}
\label{dorsum_area_sampleB}
\end{figure}

\section{Implementation}
\label{implementation}

This section introduces the dataset used for data training and evaluation of the CapillaryNet architecture.
Moreover, the annotation of the data and the implementation of the system are also described.

\subsection{Dataset}

The videos are captured on human subjects using a handheld digital microscope (Digital Capillaroscope, Inspectis, Sweden) at a resolution of 1920 × 1080 at 30 frames per second for 20 seconds.
The nutritive capillaries in the skin papillae in the dorsum region of the hand are visualised in the videos, as depicted in Figure \ref{dorsum_area_sampleA}.
A sample of a video frame is shown in Figure \ref{dorsum_area_sampleB}.
The region is coated with a layer of transparent oil before applying the microscope to the skin.
The total dataset consists of 365 videos captured from 50 subjects.
Each video in the dataset depicts at least one capillary, with some depicting up to 70 capillaries.
All subjects provided informed consent to participate in the study (IRB protocol and approval number: 2020P001987). The average age of the subjects is 57 years with a standard deviation of 17 years.

\subsection{Data Annotation}

A trained researcher annotated the obtained microvascular videos using in-house software. Capillaries visible in different frames of each video are marked with bounding boxes and assigned velocity labels whenever applicable. For capillary detection training, a single frame is extracted from the bounding boxes, and for velocity detection, multiple frames are extracted from within the bounding boxes. The two stated steps are performed to obtain the training, validation, and test datasets. A second trained researcher labelled the dataset, and the reproducibility of capillary detection between two researchers is calculated to be $\sim$84\%. Thus, the accuracy of the trained researcher is assumed to be $\sim$84\%. Our industrial stakeholder, ODI Medical AS (a MedTech company specializing in microcirculation analysis over the past decade), confirmed this value.

The capillary detection algorithm is trained on $\sim$70\% of the dataset, validated on $\sim$20\%, and tested on $\sim$10\%.
The total dataset consists of 365 videos with $\sim$12,000 capillary images and $\sim$2,000 non-capillary images, with a ratio of 6:1.
This class imbalance is considered during the design of the deep neural network, and the prediction accuracies are reflected in the F1 scores obtained.
The algorithm’s accuracy on capillary detection and velocity classification is calculated by comparing the labels of the bounding boxes assigned by the trained researcher with those obtained using the proposed system.

\subsection{System Implementation}

The following packages were used to implement the system
imageio [v2.13.4], 
imutils [v0.5.4],
keras [v2.7.0],
Keras-Preprocessing [v1.1.2],
matplotlib [v3.5.1],
numpy [v1.21.5],
opencv-python [v4.5.4.60] \cite{bradski2000opencv},
Pillow [v8.4.0],
scikit-image[v0.19.1],
scikit-learn [v1.0.1] \cite{pedregosa2011scikit},
scipy [v1.7.3] and
tensorflow [2.7.0] \cite{tensorflow1},
Python 3.9~\cite{van2007python}.

Coding and evaluation were performed in Pycharm Professional 2021.1 on a computer with Windows 10 operating system.
A demo of the system can found here http://www.analysecapillary.space/.
The system can be cloned, modified, and used by following the instructions provided on the website.
The user can clone the package from the Github repository and import it into their Python environment to use the system under the stated license \footnote{CapillaryNet: An Automated System to Quantify Skin Capillary Density and Red Blood Cell Velocity from Handheld Vital Microscopy © 2020 by Maged Abdalla Helmy Mohamed Abdou is licensed under Attribution-NonCommercial 4.0 International. To view a copy of this license, visit http://creativecommons.org/licenses/by-nc/4.0/}.

\section{Results and Discussion}
\label{resultsandDiscussion}

In this section, the results obtained by using CapillaryNet are presented and discussed.
CapillaryNet is compared with other object detection algorithms, as described in Section \ref{method_accuracy}.
CapillaryNet is compared with other microcirculation analysis methods in the literature, as detailed in Section \ref{method_accuracy_2}. Sections
\ref{method_accuracy_3} and 
\ref{cap_net_speed} 
provide a discussion on the first stage results (capillary hematocrit and capillary density) and present the accuracy of the proposed method.
Sections
\ref{method_accuracy_4},
\ref{method_accuracy_5} and
\ref{method_accuracy_6}, 
present the second stage (flow velocity of blood within capillaries and their direction) results.
Sections
\ref{method_accuracy_8},
\ref{method_accuracy_10},
\ref{method_accuracy_11},
and
\ref{method_accuracy_15}
present an ablation study, generalisation of the proposed framework, the limitations of this paper, and future research directions.

\subsection{Capillary Detection: Comparison of CapillaryNet with other Object Detection Algorithms}
\label{method_accuracy}

Object detection is one of the most challenging problems in computer vision.
The aim of this task is to detect the area occupied by an object in an image \cite{zhao2019object}. Deep learning is a powerful tool, which significantly increased the mean average precision in various object detection competitions (i.e., VOC 2007-2012 and ILSVRC 2013-2017) \cite{krizhevsky2012imagenet,russakovsky2015imagenet}.
Before deep learning, salient detection methods were state-of-the-art in object detection \cite{wang2021salient,borji2015salient,borji2019salient}.

Object detection algorithms involve millions of parameters to optimise with respect to specific datasets \cite{zhao2019object}.
For example, the widely used Mask R-CNN algorithm achieved state-of-the-art object segmentation results on the COCO dataset (comprising 80 object categories with 1.5 million object instances) \cite{zhao2019object,lin2014microsoft}.
Mask R-CNN is not suitable for use in tightly constrained environments for several reasons. First, Mask R-CNN involves $\sim$6.3 million optimisation parameters, and our experiments show that training it on the same dataset and hardware as CapillaryNet requires approximately five days.
In contrast, CapillaryNet exhibits a training time of $\sim$20 seconds per epoch , as its bounding box detection algorithm only requires $\sim$16,860 parameters.
Moreover, the deployment of large models requires relatively high-end computers that utilise GPUs that are difficult to find in medical devices.
Thus, CapillaryNet was designed to operate in medical environments with limited computing capabilities while simultaneously exceeding the accuracy of a trained researcher.
This is enabled by the fact that CapillaryNet utilises a much shallower convolutional neural network (CNN) compared to object detection algorithms such as Mask R-CNN.
Table \ref{tab:overview} lists some commonly used models and the corresponding numbers of optimisation parameters \cite{ai4everyone}— DenseNet201 requires 18.28 million parameters, whereas ResNet152 requires 58.34 million parameters.
Thus, a shallower CNN makes the proposed model more versatile by enabling faster re-training when new data is incorporated.
With a training speed that is faster by a factor of 1000, a detection time that is better by a factor of 100, and a fraction of the size of the current state-of-the-art object detection algorithms, CapillaryNet is a production-ready capillary detection model that is optimised to perform microcirculation analysis in clinical environments.

\begin{figure}[!ht]
\center
\includegraphics[width=0.9\linewidth]{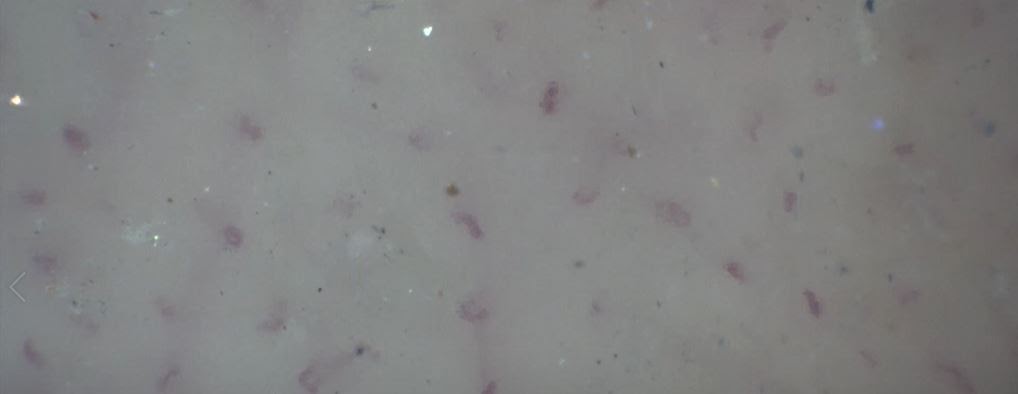}
\caption{Image of the capillaries on the dorsum part of the hand with dirt and microscopic artifacts. Using purely rule-based algorithms to detect capillaries in this case will lead to the detection of the artifacts due to similarities in size. On the other hand, by combining salient detection methods with convolutional neural networks, the algorithms can distinguish between capillaries and artifacts}
\label{artificats_on_skin}
\end{figure}

Furthermore, the experiments conducted in this paper indicate that CNN-based RoI detectors, e.g., that of Mask R-CNN, do not generalise well with new capillary data.
This can be attributed to the variation in illumination over the image, the blurring induced by the size of a capillary relative to that of the image, or occlusion of the skin by hair, stains, and other artifacts, as indicated in Figure \ref{artificats_on_skin}.
The CapillaryNet architecture tackles these challenges posed by the skin profile by using the HSV colour space to detect RoIs instead of pure CNNs. Moreover, the RoI detection methods utilised by CapillaryNet are computationally less expensive than that of Mask R-CNN.

\begin{figure}[!ht]
\center
\begin{tabular}{l}
\includegraphics[width=0.9\linewidth]{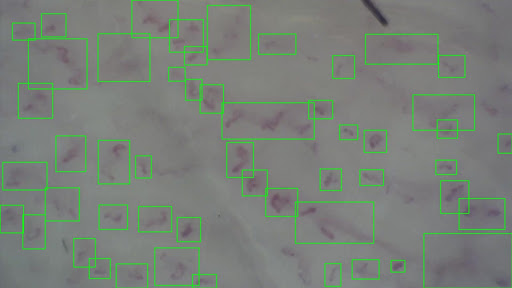}                   \\
\multicolumn{1}{c}{a} \\
\includegraphics[width=0.9\linewidth]{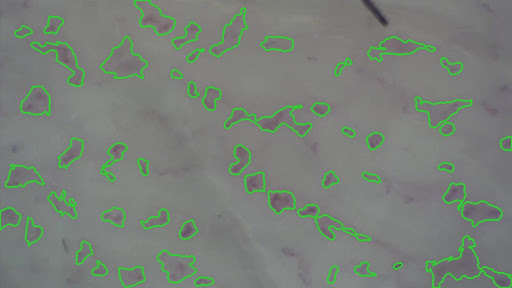}                     \\
\multicolumn{1}{c}{b} \\
\includegraphics[width=0.9\linewidth]{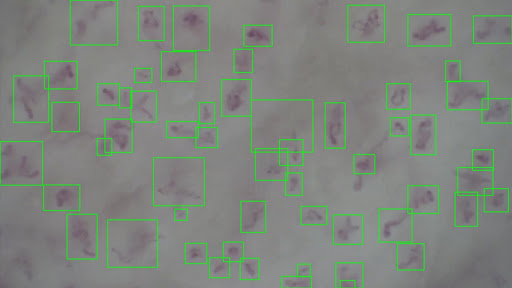}                   \\
\multicolumn{1}{c}{c} \\
\includegraphics[width=0.9\linewidth]{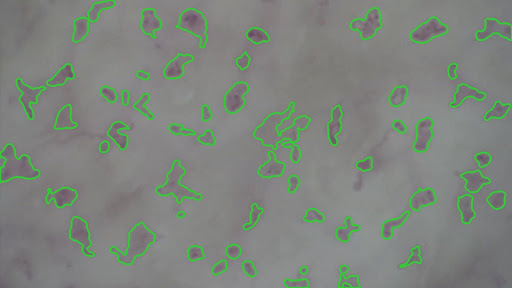}                    \\
\multicolumn{1}{c}{d}
\end{tabular}
\caption{Capillary detection using CapillaryNet. Areas with capillaries are marked in green. The bounding boxes are shown in images a and images c. Images b and images d depicts the corresponding results of applying the mask}
\label{boundingBox_output}
\end{figure}

The bounding boxes obtained by using CapillaryNet are illustrated in Figure \ref{boundingBox_output}a.
Figure \ref{boundingBox_output}b highlights the area occupied by a capillary.
Similarly, Figures \ref{boundingBox_output}c and \ref{boundingBox_output}d depict another image obtained by using a handheld microscope and the area occupied by capillaries, respectively. As the lines are tightly wrapped around the capillaries within the bounding box, the area occupied can be calculated with high accuracy, and the capillary density is derived from that value.

\begin{table}[]
\caption{Comparison of the size, number of parameters, and training time of CapillaryNet with those of other commonly used object detection algorithms}
\label{tab:overview}
\resizebox{\columnwidth}{!}{%
\begin{tabular}{|c|c|c|c|}
\hline
\textbf{Model Name} & \textbf{Size (MB)} & \textbf{Param (Mil)} & \textbf{training time} \\ \hline
CapillaryNet        & $\sim$10           & $\sim$0.6      & $\sim$minutes  \\ \hline
Mask RCNN           & $\sim$256          & $\sim$6.3      & $\sim$days   \\ \hline
DenseNet201         & $\sim$7947         & $\sim$18.28    & $\sim$weeks   \\ \hline
ResNet152           & $\sim$9097         & $\sim$58.34    & $\sim$weeks   \\ \hline
\end{tabular}%
}
\end{table}

\subsection{Capillary Detection: Comparison of CapillaryNet with Other Capillary Detection Methods Reported in the Literature}
\label{method_accuracy_2}

CapillaryNet compensates for unequal illumination using two approaches. Firstly, the application of the HSV colour space enhances the robustness of the architecture against changes in illumination.
Secondly, the inclusion of out-of-focus capillaries in training, validation, and test datasets, followed by training of the CNN on the dataset labelled by a trained researcher, further reduces observer variation.
Moreover, several methods in the literature highlighted the need for averaging the pixel values of several frames. In contrast, the utilization of the HSV colour space combined with a CNN network enables CapillaryNet to detect capillaries from a single frame. Thus, averaging pixel values over the entire video and treating each frame uniquely become redundant. It also enables better real-time detection.

\subsection{Capillary Hematocrit}
\label{method_accuracy_3}

Capillary hematocrit is indicative of the potential of each capillary to deliver oxygen to its surrounding tissues.
When capillaries have a regular flow and are normally distributed but exhibit a low concentration of red blood cells (i.e., low hematocrit), the efficiency of oxygen delivery via microcirculation may be compromised.
Additional clinical studies are required to elucidate the potential of using capillary hematocrit as an indicator for abnormal microcirculation.
The proposed method calculates capillary hematocrit.

\begin{figure}[!ht]
\begin{tabular}{cc}
\includegraphics[width=0.4\columnwidth]{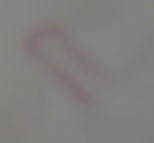}         & 
\includegraphics[width=0.4\columnwidth]{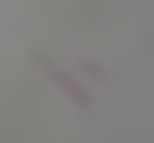}         \\
a         & b         \\
\multicolumn{2}{c}{\includegraphics[width=0.4\columnwidth]{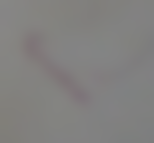}}   \\
\multicolumn{2}{c}{c} \\
\end{tabular}
\caption{a) presents a capillary filled with red blood cells. (b) and (c) display the same capillary after a few seconds with white blood cells and plasma gaps.
Figure \ref{Capillary_Hematocrit_graph} shows the red blood cell distribution of this capillary derived using CapillaryNet}
\label{Capillary_Hematocrit}
\end{figure}

\begin{figure}[!ht]
\begin{tabular}{c}
\includegraphics[width=0.9\columnwidth]{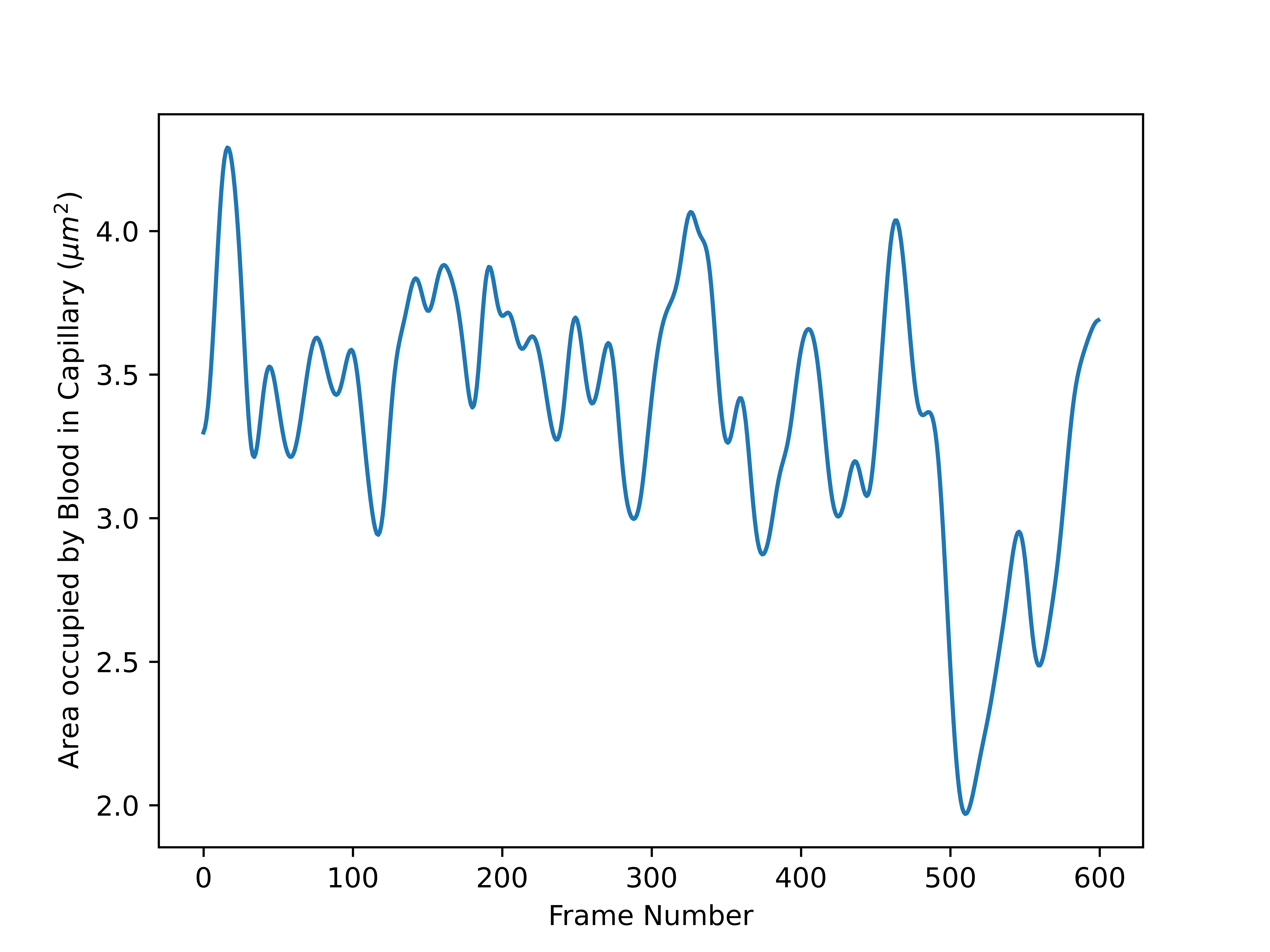} \\
a
\end{tabular}
\caption{Graph representing the proportion of a capillary filled with red blood cells (capillary hematocrit) with respect to time)
}
\label{Capillary_Hematocrit_graph}
\end{figure}

Figure \ref{Capillary_Hematocrit}a presents a capillary filled with red blood cells. Figures \ref{Capillary_Hematocrit}b and \ref{Capillary_Hematocrit}c display the same capillary after a few seconds with white blood cells and plasma gaps. Figure \ref{Capillary_Hematocrit_graph} shows the red blood cell distribution of this capillary derived using CapillaryNet.

\subsection{CapillaryNet Speed and Accuracy}
\label{cap_net_speed}

CapillaryNet can analyse 30 frames in $\sim$4 s , which is the equivalent of a speed of $\sim$0.9 s per frame, whereas a trained analyst spends 2 min on average on the same frame to highlight the capillaries and calculate capillary density.
As mentioned in the Implementation section, the accuracy of a trained researcher is $\sim$84\%, whereas that of CapillaryNet is $\sim$93\%.
Table \ref{tab:table_1} summarises the main differences between a trained researcher and CapillaryNet in terms of capillary detection. Furthermore, it shows the F1 scores of other state of art models.
The macroscopic average accuracy on the test dataset of the HSV CNN model is $\sim$96\%, and that on the SSIM CNN model is $\sim$86\%.
Table \ref{models_HSV_F1_score} summarises the F1 scores of HSV Model and Table \ref{models_SSIM_F1_score} summarises the F1 scores of SSIM Model.
The outputs obtained from both CNNs are combined to maximise capillary detection in a microcirculation image.

\begin{table}[!ht]
\caption{Benchmark of CapillaryNet capillary detection against manual analysis performed by a trained researcher on a microcirculation video and other state of the art methods}
\bigskip
\label{tab:table_1}
\resizebox{\columnwidth}{!}{%
\begin{tabular}{|c|c|c|c|}
\hline
\textbf{Name}               & \textbf{Training Time} & \textbf{Detection Time} & \textbf{F1 Score} \\ \hline
\textbf{DenseNet121 - Backbone}        & $\sim$5 minutes         & $\sim$3 seconds       & $\sim$70\%        \\ \hline
\textbf{ResNet152  - Backbone}           & $\sim$7 minutes         & $\sim$5 seconds       & $\sim$73\%        \\ \hline
\textbf{Trained Researcher} & $\sim$80 hours         & $\sim$120 seconds       & $\sim$84\%        \\ \hline
\textbf{Mask RCNN}          & $\sim$30 hours         & $\sim$100 seconds       & $\sim$86\%        \\ \hline
\textbf{CapillaryNet}       & $\sim$2 minutes        & $\sim$0.9 seconds       & $\sim$93\%        \\ \hline
\end{tabular}%
}
\end{table}

\begin{table}[!ht]
\caption{F1-scores of the HSV CNN Model on the validation and test sets}
\bigskip
\label{models_HSV_F1_score}
\resizebox{\columnwidth}{!}{%
\begin{tabular}{|ccccc|}
\hline
\multicolumn{5}{|c|}{\textbf{Accuracy on Validation Dataset}} \\ \hline
\multicolumn{1}{|c|}{\cellcolor[HTML]{000000}\textbf{}} &
  \multicolumn{1}{c|}{\textbf{Precision}} &
  \multicolumn{1}{c|}{\textbf{Recall}} &
  \multicolumn{1}{c|}{\textbf{F1-score}} &
  \textbf{Number of  Samples} \\ \hline
\multicolumn{1}{|c|}{\textbf{Capillary}} &
  \multicolumn{1}{c|}{0.99} &
  \multicolumn{1}{c|}{0.98} &
  \multicolumn{1}{c|}{0.99} &
  2781 \\ \hline
\multicolumn{1}{|c|}{\textbf{Non Capillary}} &
  \multicolumn{1}{c|}{0.91} &
  \multicolumn{1}{c|}{0.94} &
  \multicolumn{1}{c|}{0.93} &
  536 \\ \hline
\multicolumn{1}{|c|}{\textbf{Accuracy}} &
  \multicolumn{2}{c|}{\cellcolor[HTML]{000000}} &
  \multicolumn{1}{c|}{0.98} &
  3317 \\ \hline
\multicolumn{1}{|c|}{\textbf{Macro avg}} &
  \multicolumn{1}{c|}{0.95} &
  \multicolumn{1}{c|}{0.96} &
  \multicolumn{1}{c|}{0.96} &
  3317 \\ \hline
\multicolumn{1}{|c|}{\textbf{Weighted avg}} &
  \multicolumn{1}{c|}{0.98} &
  \multicolumn{1}{c|}{0.98} &
  \multicolumn{1}{c|}{0.98} &
  3317 \\ \hline
\multicolumn{5}{|c|}{\textbf{Accuracy on Test Dataset}} \\ \hline
\multicolumn{1}{|c|}{\cellcolor[HTML]{000000}\textbf{}} &
  \multicolumn{1}{c|}{\textbf{Precision}} &
  \multicolumn{1}{c|}{\textbf{Recall}} &
  \multicolumn{1}{c|}{\textbf{F1-score}} &
  \textbf{Number of  Samples} \\ \hline
\multicolumn{1}{|c|}{\textbf{Capillary}} &
  \multicolumn{1}{c|}{0.96} &
  \multicolumn{1}{c|}{0.97} &
  \multicolumn{1}{c|}{0.96} &
  918 \\ \hline
\multicolumn{1}{|c|}{\textbf{Non Capillary}} &
  \multicolumn{1}{c|}{0.84} &
  \multicolumn{1}{c|}{0.78} &
  \multicolumn{1}{c|}{0.81} &
  187 \\ \hline
\multicolumn{1}{|c|}{\textbf{Accuracy}} &
  \multicolumn{2}{c|}{\cellcolor[HTML]{000000}} &
  \multicolumn{1}{c|}{0.94} &
  1105 \\ \hline
\multicolumn{1}{|c|}{\textbf{Macro avg}} &
  \multicolumn{1}{c|}{0.90} &
  \multicolumn{1}{c|}{0.88} &
  \multicolumn{1}{c|}{0.89} &
  1105 \\ \hline
\multicolumn{1}{|c|}{\textbf{Weighted avg}} &
  \multicolumn{1}{c|}{0.94} &
  \multicolumn{1}{c|}{0.94} &
  \multicolumn{1}{c|}{0.94} &
  1105 \\ \hline
\end{tabular}%
}
\end{table}

\begin{table}[!ht]
\caption{F1-scores of the SSIM CNN Model on the validation and test sets}
\bigskip
\label{models_SSIM_F1_score}
\resizebox{\columnwidth}{!}{%
\begin{tabular}{|ccccc|}
\hline
\multicolumn{5}{|c|}{\textbf{Accuracy on Validation Dataset}} \\ \hline
\multicolumn{1}{|c|}{\cellcolor[HTML]{000000}\textbf{}} &
  \multicolumn{1}{c|}{\textbf{Precision}} &
  \multicolumn{1}{c|}{\textbf{Recall}} &
  \multicolumn{1}{c|}{\textbf{F1-score}} &
  \textbf{Number of  Samples} \\ \hline
\multicolumn{1}{|c|}{\textbf{Capillary}} &
  \multicolumn{1}{c|}{0.98} &
  \multicolumn{1}{c|}{0.99} &
  \multicolumn{1}{c|}{0.99} &
  969 \\ \hline
\multicolumn{1}{|c|}{\textbf{Non Capillary}} &
  \multicolumn{1}{c|}{0.86} &
  \multicolumn{1}{c|}{0.57} &
  \multicolumn{1}{c|}{0.69} &
  56 \\ \hline
\multicolumn{1}{|c|}{\textbf{Accuracy}} &
  \multicolumn{2}{c|}{\cellcolor[HTML]{000000}} &
  \multicolumn{1}{c|}{0.97} &
  1025 \\ \hline
\multicolumn{1}{|c|}{\textbf{Macro avg}} &
  \multicolumn{1}{c|}{0.92} &
  \multicolumn{1}{c|}{0.78} &
  \multicolumn{1}{c|}{0.84} &
  1025 \\ \hline
\multicolumn{1}{|c|}{\textbf{Weighted avg}} &
  \multicolumn{1}{c|}{0.97} &
  \multicolumn{1}{c|}{0.97} &
  \multicolumn{1}{c|}{0.97} &
  1025 \\ \hline
\multicolumn{5}{|c|}{\textbf{Accuracy on Test Dataset}} \\ \hline
\multicolumn{1}{|c|}{\cellcolor[HTML]{000000}\textbf{}} &
  \multicolumn{1}{c|}{\textbf{Precision}} &
  \multicolumn{1}{c|}{\textbf{Recall}} &
  \multicolumn{1}{c|}{\textbf{F1-score}} &
  \textbf{Number of  Samples} \\ \hline
\multicolumn{1}{|c|}{\textbf{Capillary}} &
  \multicolumn{1}{c|}{0.97} &
  \multicolumn{1}{c|}{0.98} &
  \multicolumn{1}{c|}{0.98} &
  595 \\ \hline
\multicolumn{1}{|c|}{\textbf{Non Capillary}} &
  \multicolumn{1}{c|}{0.80} &
  \multicolumn{1}{c|}{0.68} &
  \multicolumn{1}{c|}{0.74} &
  57 \\ \hline
\multicolumn{1}{|c|}{\textbf{Accuracy}} &
  \multicolumn{2}{c|}{\cellcolor[HTML]{000000}} &
  \multicolumn{1}{c|}{0.96} &
  652 \\ \hline
\multicolumn{1}{|c|}{\textbf{Macro avg}} &
  \multicolumn{1}{c|}{0.88} &
  \multicolumn{1}{c|}{0.83} &
  \multicolumn{1}{c|}{0.86} &
  652 \\ \hline
\multicolumn{1}{|c|}{\textbf{Weighted avg}} &
  \multicolumn{1}{c|}{0.95} &
  \multicolumn{1}{c|}{0.96} &
  \multicolumn{1}{c|}{0.96} &
  652 \\ \hline

\end{tabular}%
}
\end{table}

\subsection{Velocity Detection}
\label{method_accuracy_4}

The quantification of red blood cell flow is a more challenging task than capillary detection. Some papers have described red blood cell flow based on manual quantification of red blood cells of different scales \cite{boerma2005quantifying,arnold2009point,edul2012quantitative,dubin2009increasing}, which is subject to intra-individual variation. In manual quantification, each individual vessel is assigned a score representing the average flow velocity estimated by a researcher. Flow velocity scales vary between publications—some researchers have classified flow on a scale of 0–2 (absent, intermittent, or continuous flow) \cite{de2007evaluate}, others on a scale of 0–3 (absent, intermittent, sluggish, or normal flow) \cite{de2007evaluate}, while others on a scale of 0–5 (no-flow, sluggish, continuous very low, continuous low, continuous high, or brisk flow) \cite{fredly2016noninvasive}.

More recent papers have used space-time diagrams (STDs) \cite{hilty2019microtools,bezemer2011rapid,hilty2017assessment} to quantify red blood cell flow, which exhibit a fundamental improvement over manual analysis as STDs are independent of intra-individual variation \cite{edul2012quantitative, dobbe2008measurement}.
However, STDs are susceptible to slight movement due to their reliance on the position of the central line along the centre of the capillary in all frames. Thus, any movement necessitates additional manual recalibration, increasing the risk of errors and user bias. Moreover, constructing an accurate central line depends on the accurate identification of the exact width and length of the capillaries at an earlier stage. Therefore, out-of-focus capillaries need to be disregarded while fitting the central line. Moreover, STDs can only be constructed based on the whole video.

CapillaryNet provides an alternative to STDs and manual analysis, which is considered to be the gold standard for red blood cell (RBC) velocity classification \cite{hilty2019microtools}. It does not require a central line to deduce the velocity; which reduces the number of steps required. Further, the proposed system can be used to deduce intra-capillary flow heterogeneity. This enhances the accuracy of tracking RBCs of the proposed method.

CapillaryNet method can be considered as an alternative way to the 
space-time diagram and manual eye analysis, which is considered as the gold standard 
for red blood cells (RBC) velocity classification \cite{hilty2019microtools}. The 
central line is not needed to deduce the velocity; therefore, the number of steps are reduced by 1 in comparison to the gold standard of detecting velocity.
Furthermore, the proposed system can be used to deduce the intra-capillary 
flow heterogeneity. This makes the proposed method more accurate in tracking red blood 
cells the reliance on a central line is accurately placed between 
the frames. 

\begin{figure}[!ht]
\center
\includegraphics[width=\linewidth]{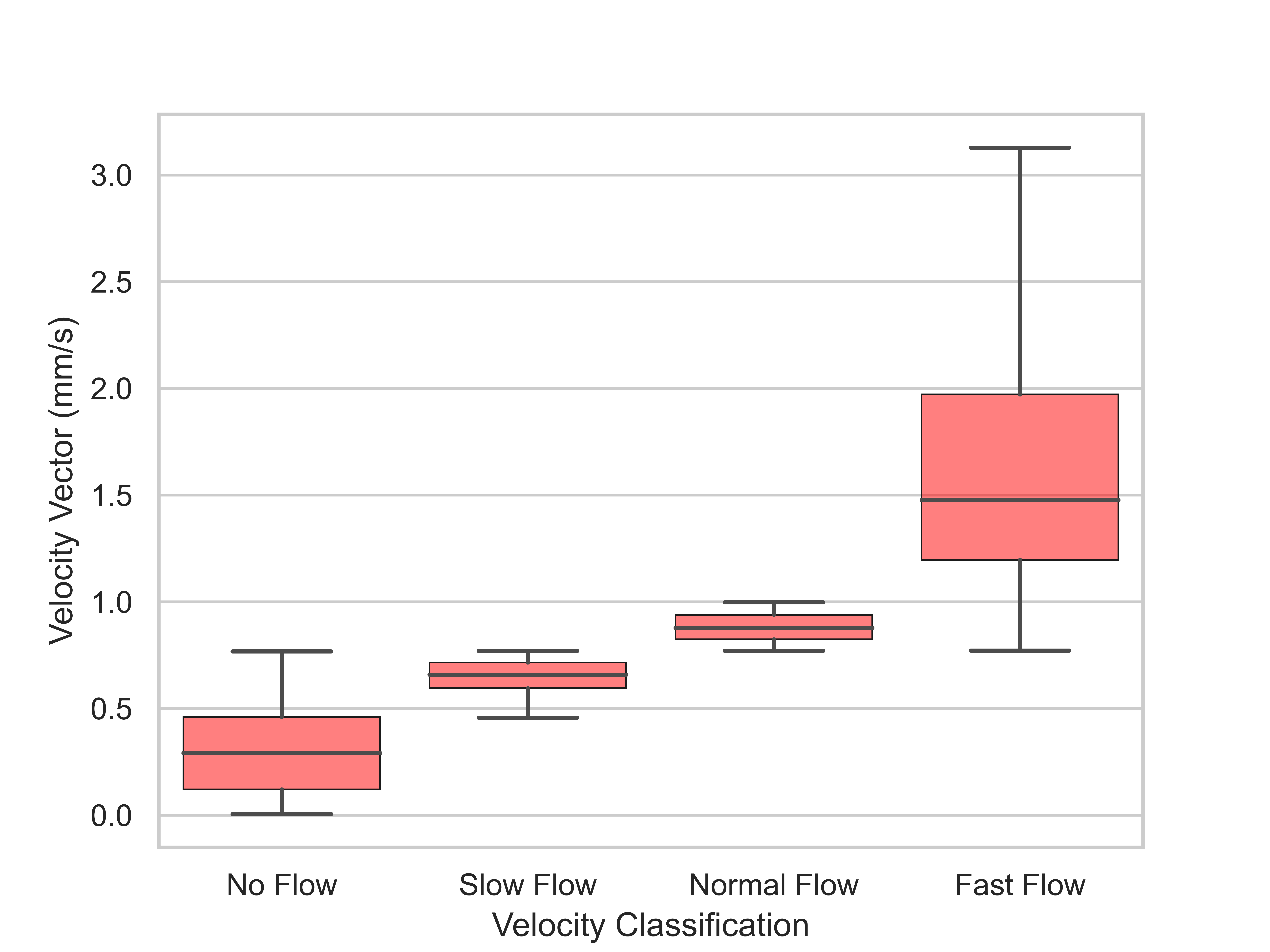}
\caption{
Velocity vector values of 2,255 capillary videos classified by CapillaryNet. The average velocity vector for no flow is $\sim$0.3 mm/s, that for slow flow is $\sim$0.63 mm/s, that for normal flow is $\sim$0.88 mm/s, and that for fast flow is $\sim$1.66 mm/s
}
\label{GF_algorithm}
\end{figure}

To evaluate the proposed method, a trained researcher classified the velocities of about 2250 capillaries based on a sequence of 240 frames (8 seconds) captured in high resolution (1920 × 1080) at 30 frames per second (fps) into the following categories: no-flow, slow flow, normal flow, and fast flow. For each capillary classified by the analyst, CapillaryNet yields a velocity vector. This value represents the rate of change in capillary position, which is equated with speed. Figure \ref{GF_algorithm} depicts a comparison of the analyst’s classification (depicted on the X-axis) and the velocity vectors obtained using the proposed algorithm (Y-axis). The velocity vectors are observed to be significantly different for each velocity class (the p-values between the pairs of classes are below 0.00001). The magnitude of the average velocity vector  for no-flow is  $\sim$0.3 mm/s, that for slow flow is  $\sim$0.63 mm/s, that for normal flow is  $\sim$0.88 mm/s, and that for fast flow is  $\sim$1.66 mm/s, as recorded in Figure \ref{GF_algorithm} and Table \ref{GF_algorithm_table}.

The overlap in velocity vector values between the velocity classes, no-flow and slow flow, and the classes, normal flow and fast flow, indicates the lack of precise definition of the boundaries between velocity categories in manual analysis. For this reason, analysts may classify a capillary with borderline fast-normal flow into the normal flow category or the fast flow category. Hubble et al. \cite{hubble2009variability} demonstrated that inter-analyst variability can extend to up to 26\% in the assessment of capillary flow in healthy controls. The accuracy is also dependent on the experience of the analyst \cite{fredly2015skin}. This lack of accuracy in borderline cases creates inconsistent training data for the algorithm and results in overlapping velocity vector values between categories.

The capillaries classified by the analysts under fast flow exhibited a broader range of velocity vectors compared to the other velocity classes, as shown in Figure \ref{GF_algorithm}. This indicates that the algorithm successfully splits the fast flow category into two categories (e.g., fast and swift flow). This novel classification is expected to provide insights into the variation of capillary flow velocity during the course of various diseases.

The algorithm is trained on the classification of a single analyst and the velocity vector boundaries between categories are refined to create a standardised method for enhancing classification consistency for capillary velocities. This reduces the intra-analyst variation in velocity classification and is not influenced by parameters such as experience of concentration of the analyst. Further, the GF algorithm can detect flow in near real-time at almost 15 fps ($\sim$0.07 s per bounding box), whereas an analyst spends 1 min on average to label the speed of a single capillary and 20 min on average to annotate a single video.

\begin{table}[]
\caption{Velocity classification and the velocity vector of the dataset}
\bigskip
\label{GF_algorithm_table}
\resizebox{\columnwidth}{!}{%
\begin{tabular}{|c|c|c|}
\hline
\textbf{Velocity Classification} & \textbf{\begin{tabular}[c]{@{}c@{}}Velocity Vector \\ (mm/s)\end{tabular}} & \textbf{Accuracy} \\ \hline
No Flow     & $\sim$0.30 & $\sim$72\% \\ \hline
Slow Flow   & $\sim$0.63 & $\sim$80\% \\ \hline
Normal Flow & $\sim$0.88 & $\sim$85\% \\ \hline
Fast Flow   & $\sim$1.66 & $\sim$90\% \\ \hline
\end{tabular}%
}
\end{table}

\subsection{Deep Neural Networks for Velocity Classification}
\label{method_accuracy_5}

\begin{figure}[!ht]
\center
\includegraphics[width=0.9\columnwidth]{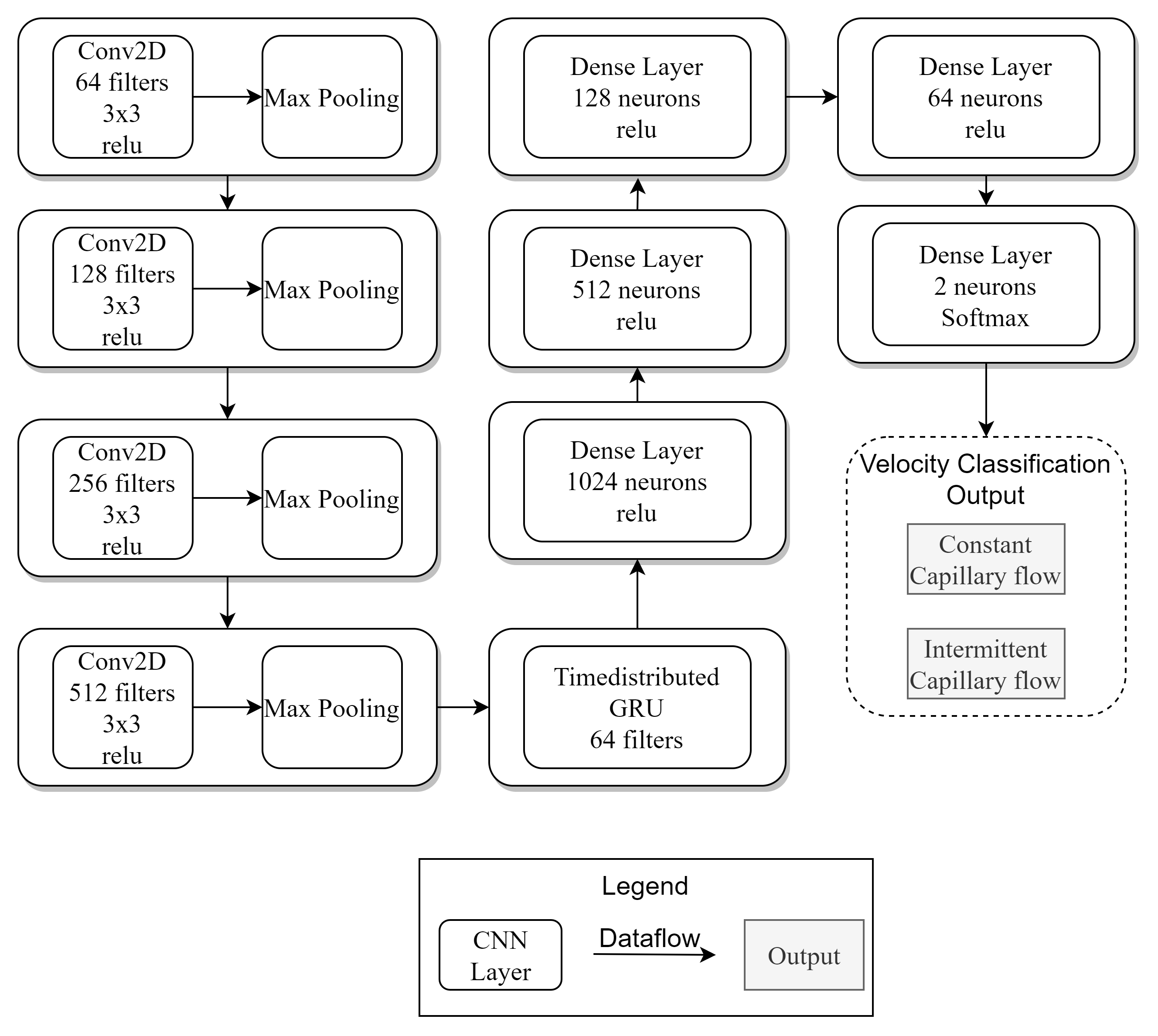}
\caption{A deep neural architecture built to evaluate velocity classification}
\label{velocityDetection_DNN}
\end{figure}

Classification of frames at a resolution of 1920 × 1080 at 30 fps requires a deep learning model whose
application in a clinical environment is not feasible. Therefore, experiments are performed at a lower resolution (1280 × 720) and lower fps (5fps) for velocity classification.
This section presents four different deep neural architectures designed to train velocity classification based on videos. This architecture is trained and developed to evaluate the velocity classification performance of CapillaryNet.
The presented architectures are based on state-of-the-art papers \cite{szegedy2017inception,deng2009imagenet,sherstinsky2020fundamentals,maturana2015voxnet}
In the first experimental architecture, InceptionResNetV2~\cite{szegedy2017inception} is used in combination with transfer learning. The top nine layers are fine-tuned on the four classes (no-flow, slow flow, normal velocity flow, and fast flow), whereas the remainder of the weights are adopted based on the ImageNet dataset~\cite{deng2009imagenet}. The output is then transmitted to a convolutional 2D network.
The second architecture is based on a time-distributed CNN, and the features of the CNN are transmitted to a recurrent neural network (RNN) \cite{sherstinsky2020fundamentals}. For the RNN network, experiments with both long short-term memory (LSTM) and gated recurrent unit (GRU) are conducted. The architecture for a time-distributed CNN equipped with an RNN for velocity classification is depicted in Figure \ref{velocityDetection_DNN}.
The third architecture uses a set of 3D convolutional networks \cite{maturana2015voxnet}. 
The fourth architecture is similar to the first architecture.
It also utilises transfer learning of InceptionResNetV2, but unlike the first case, its output is transmitted to a convolutional 3D network. Thus, the fourth architecture is a combination of the first and third architectures. All four aforementioned architectures are observed to perform poorly, with accuracies less than 50\% corresponding to each class. Moreover, these algorithms involve several million parameters to be trained, drastically increasing their energy consumption and training durations compared to CapillaryNet. Thus, classification of capillary velocity using DNNs is complex and energy-consuming, and it remains a challenging task.

\begin{figure}[!ht]
\center
\begin{tabular}{c}\\
{\includegraphics[width=0.5\columnwidth,scale=0.3]{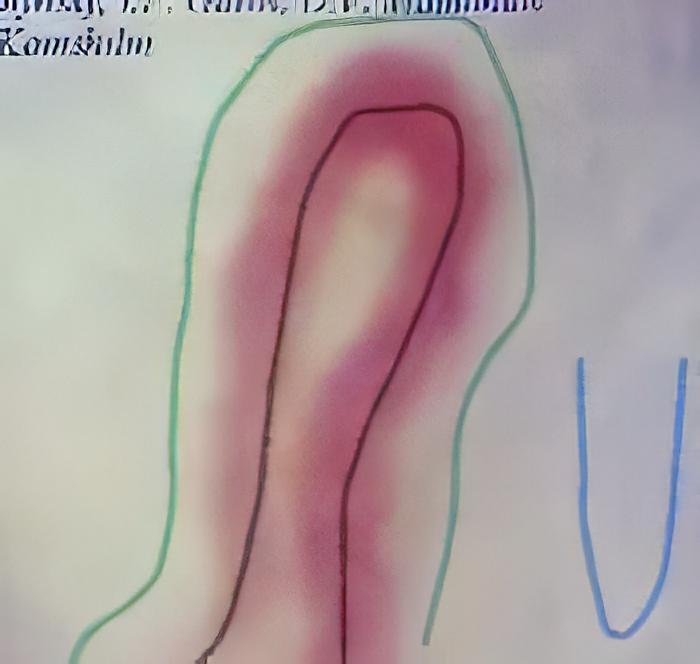}}\\
{a}                 \\
{\includegraphics[width=0.9\linewidth]{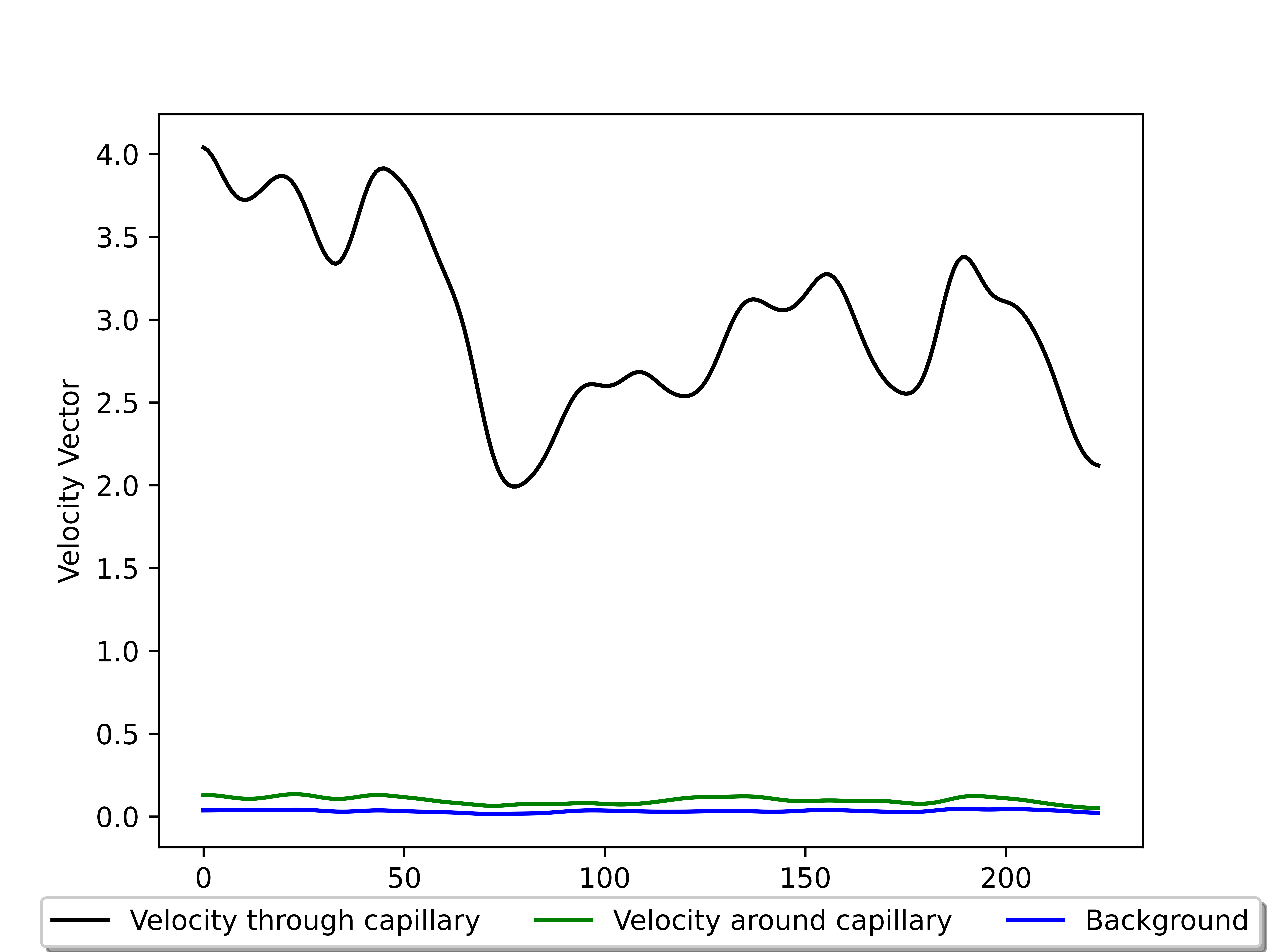}}\\
{b}        \\
\end{tabular}
\caption{(a) CapillaryNet applied to a capillary video publicly available to assess capillary flow direction and heterogeneity. Three lines were plotted in the video—the black line is approximately aligned with the centre of the capillary, the green line is placed around the capillary, and the blue line is randomly placed next to the capillary. These lines are visible in the image. (b) Results of the velocity vector across the frames of the black line (average velocity vector value is approximately 3), the green line (average velocity vector value is 0.2), and the blue line (average velocity vector value is 0.1)}
\label{intraFlow}
\end{figure}

\begin{table}[!ht]
\label{tab:hiltyOverview}
\caption{
Estimation of applicable parameters suggested for microcirculation analysis by Hilty et al \cite{hilty2019microtools} and Ince et al \cite{ince2018second}. Two new parameters can be uniquely identified and calculated using CapillaryNet—Intracapillary heterogeneity of flow velocity and capillary hematocrit}
\resizebox{\columnwidth}{!}{%
\begin{tabular}{|c|c|}
\hline
\textbf{Parameter} &
  \textbf{\begin{tabular}[c]{@{}c@{}}CapillaryNet Detection \\ Description\end{tabular}} \\ \hline
\multicolumn{2}{|c|}{\textbf{per image or video}} \\ \hline
\begin{tabular}[c]{@{}c@{}} Total vessel Density \\for Capillaries\end{tabular} &
  \begin{tabular}[c]{@{}c@{}}
  The sum of the area occupied by \\ 
  the capillaries is derived based \\ 
  on the total number of  \\ 
  pixels occupied \\ 
  by the detected capillary divided \\ 
  by the dimension of the image\end{tabular} \\ \hline

\begin{tabular}[c]{@{}c@{}}Functional Capillary \\Density\end{tabular} &
  \begin{tabular}[c]{@{}c@{}}
  Sum of the area occupied by  \\ 
  capillaries that contain moving  \\ 
  RBCs is derived based on the   \\ 
  total number of pixels \\ 
  occupied by the detected  \\ 
  capillary divided by the \\
  dimension of the image \end{tabular} \\ \hline
  
\multicolumn{2}{|c|}{\textbf{per vessel}} \\ \hline
Flow Velocity &
  \begin{tabular}[c]{@{}c@{}}
  The velocity of red blood cells \\ 
  flow classified as no flow, slow flow,\\ 
  normal velocity flow and fast flow
  \end{tabular} \\ \hline
\begin{tabular}[c]{@{}c@{}}Intra-capillary \\ heterogeneity of \\ flow velocity\end{tabular} &
  \begin{tabular}[c]{@{}c@{}}
  Derived from the velocity \\ 
  part of CapillaryNet \\ 
  where the the variation \\ 
  in pixel movement is plotted \\
  across all frames\end{tabular}
  \\ \hline
\begin{tabular}[c]{@{}c@{}}Capillary \\ Hematocrit\end{tabular} &
  \begin{tabular}[c]{@{}c@{}}
  Derived using a rule-based \\
  algorithm where the \\ 
  number of red blood cells \\ 
  across the frames are plotted
  \end{tabular} \\ \hline
\end{tabular}
}
\end{table}

\begin{figure}[!ht]
\center
\includegraphics[width=0.5\columnwidth,scale=0.3]{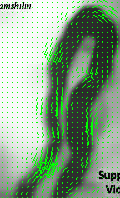}
\caption{Flow direction in a capillary automatically derived using CapillaryNet}
\label{CapillaryNetOverview_4}
\end{figure}

\subsection{Intra-capillary Flow Velocity Heterogeneity and flow direction}
\label{method_accuracy_6}

Flow velocity within a capillary has been calculated in previous studies only as an average value. Our empirical observations indicate that healthy subjects exhibit relatively homogeneous flow velocity, while in specific patient groups, the flow velocity varies significantly over the 20-second durations of the respective videos. Thus, a single average value corresponding to each capillary is incapable of capturing this difference. Therefore, measuring intra-capillary flow velocity heterogeneity can be used as an additional indicator of compromised microcirculation. CapillaryNet is capable of tracking moving pixels that represent RBCs—corresponding to bounding boxes surrounding capillaries, the pixel displacement of RBCs can be calculated. Thus the proposed method presents an alternative to STDs \cite{hilty2019microtools} and manual analysis. 

Moreover, the Intra-capillary Flow Velocity prediction accuracy of CapillaryNet is evaluated on a publicly available capillary video \cite{volkov2017video}, as depicted in Figure \ref{intraFlow}a. The measurements from three different regions of the video show that CapillaryNet can detect the movement of RBCs within capillaries. The results are recorded using a graph in Figure \ref{intraFlow}b. The highest velocity vector values are observed at the capillary centre, highlighted using the black line.
The lowest velocity is observed along the blue line, where there is no flow at all. Figure 20 highlights the flow of the RBCs within a capillary.

\subsection{CapillaryNet Ablation Study}
\label{method_accuracy_8}

Each part of the two pipelines, namely, the HSV pipeline and the SSIM pipeline, contributes to the improvement of the overall accuracy of CapillaryNet.
In this section, the effectiveness of each component in RoI generation is detailed.

In the HSV pipeline, the first step involves the application of Gaussian blur.
This is critical as it reduces the pixelation in the image and smoothens it, which reduces the number of RoIs from several thousands to several hundreds.
Next, HSV conversion is applied.
This is beneficial, as capillary detection information is spread across the entire spectra of all the RGB channels while, in the HSV colour space, most of the information is stored in saturation values lying between 60 and 255.
Further, unlike RGB, HSV separates the luma, yielding more accurate colour information corresponding to the image.
This is essential for the application of histogram equalization on a colour image, which enables the preferential increase of the intensity of capillaries compared to that of background pixels irrespective of variation in illumination or the presence of undesirable artifacts such as shadows or oil bubbles on the skin.
Enhancement of the colour and contrast further assists the HSV colour space to differentiate between capillaries and non-capillaries.
The Bitwise operator ``AND" and OTSU thresholding are mathematical operations that extract the RoI coordinates and transmit the RoIs to the CNN.
Then, the CNN filters the RoIs and outputs only those RoIs that contain a capillary. 
The first step of the SSIM Pipeline subtracts the back ground image (skin with an average pixel value of 127) from the current frame.
This generates as many as 200 RoIs.
As in the previous step, the OTSU threshold is applied to extracts the RoI coordinates and transmit the RoIs to the CNN.
Finally, the CNN filters the RoIs and outputs only those RoIs that are predicted to contain capillaries in accordance with the accuracy and parameters discussed in Section \ref{cap_net_speed}.

\subsection{Additional Benefit: More Parameters with CapillaryNet}
\label{method_accuracy_10}

Table \ref{tab:hiltyOverview} lists the estimated values of the microvascular parameters suggested by Hilty \cite{hilty2019microtools} and Ince \cite{ince2018second} obtained using CapillaryNet.
In addition, two new parameters that have not been previously monitored in microvascular videos—capillary hematocrit and intra-capillary flow velocity heterogeneity—are also monitored. Thus, the architecture of CapillaryNet is a unique combination of DNNs with salient object detection algorithms and two-frame motion estimation techniques. It is expected to motivate the development of a unified automated method for near real-time bedside analysis of microcirculation.

\subsection{System Generalization}
\label{method_accuracy_11}

The system functions and classes of CapillaryNet are designed following the design patterns of microservices.
This design philosophy allows a user of the system to replace any pipeline, phase, or stage from the proposed system with their desired methodology.
For example, the deep learning module in CapillaryNet can be replaced simply by pointing the function in the CapillaryNet code to another algorithm.
Further instructions are available in the README file in the GitHub repository.
In other cases, CapillaryNet is expected to exhibit similar accuracy, irrespective of the recording device as long as the magnification and image quality is retained.

\subsection{Limitation and Future Work}
\label{method_accuracy_15}

The velocity detection performance of the proposed system has certain limitations.
When the flow of RBCs is not perpendicular to the recording direction of the camera, the displacement of pixels across successive frames cannot be accurately measured.
Moreover, the presence of white blood cells or plasma gaps within the capillaries is essential to the accurate tracking of the flow. The authors intend to resolve this issue in future work.
Another direction is to explore extraction of information using Temporal Convolutional Networks (TCNs) \cite{xu2018end}.
TCNs operate using low-level computing features by encoding spatio-temporal information and transmitting these features to an RNN, which deduces the corresponding high-level temporal information.
Because of the distinct difference in temporal information between capillaries and the background, this method can reduce the number of RoIs generated, thereby accelerating the proposed system and increasing its overall accuracy.

\section{Conclusion}
\label{conclusion}

This paper presents a state-of-the-art automated method for microcirculation analysis.
The proposed method enables the monitoring of microvascular parameters that cannot be monitored using existing systems.
These novel parameters can be indicative of the capability of each capillary to deliver oxygen to its surrounding tissues.
Moreover, the proposed method standardises microcirculation analysis by eliminating intra-analyst variability and reducing the required duration.
In this study, CapillaryNet is used to assess skin microcirculation in the dorsum area.
CapillaryNet can automatically quantify the area occupied by a capillary, calculate the capillary density, determine the average flow velocity and the intra-capillary flow velocity heterogeneity, and quantify the capillary hematocrit.
The capillary detection speed is observed to be $\sim$0.9 seconds per frame, and its accuracy is $\sim$93\%.
Thus, CapillaryNet reduces the required duration of microcirculation analysis from 20 min to a few seconds by combining DNNs with traditional computer vision techniques.
Because of the design, the proposed system is suitable for low-power computers in clinical environments.
This is because the neural networks can be run on a CPU and the number of neural networks is minimised, while ensuring satisfactory accuracy.
The proposed system brings the state-of-the-art much closer to real-time microcirculation analysis in a clinical environment.
This claim is further reinforced by the use of the proposed system in a product in an industrial environment, where it is used to calculate and track capillary changes in patients with pancreatitis, COVID-19, and acute heart diseases.

\section{Disclosure Concerning Competing Interests}

The authors declare the following financial interests/personal interests that may be considered as
potential competing interests: 
The Research Council of Norway provided 50\% of the funds for Maged Helmy's Industrial Ph.D. (first Author - Industrial Ph.D. project nr: 305716.). His other 50\% is funded by ODI Medical AS.
Anastasiya Dykyy is 100\% funded by ODI Medical AS under BIA project nr: 282213.
ODI Medical AS has provided the hardware, funds, and expertise to collect and analyze the medical data.

\section{Research Funding}
The Research Council of Norway provided the necessary funds for this project: Industrial Ph.D.
project nr: 305716 and BIA project nr: 282213.

\section{Code}
A demo of the system can be found here http://www.analysecapillary.space/.
All information related to the code link will be available on the website.

\bibliographystyle{elsarticle-num} 
\bibliography{cas-refs}

\end{document}